\documentclass[12pt,preprint]{aastex}
\usepackage[]{apjfonts}
\usepackage{graphicx}
\usepackage{rotate}
\usepackage{times}
\usepackage[usenames,dvips]{color}
\usepackage{longtable}

\def\alphaox{$\alpha_{\rm ox}$}

\def\delchisq{$\Delta\chi^2$}
\def\feii{\ion{Fe}{2}}
\def\gams{$\Gamma_{\rm s}$}
\def\gamh{$\Gamma_{\rm h}$}
\def\gnh{$N_{\rm H}^{\rm Gal}$}
\def\hb{H$\beta$}

\def\mbh{$M_{\rm BH}$}
\def\msun{\ensuremath{M_{\odot}}}
\def\nh{$N_{\rm H}$}
\def\oiii{[\ion{O}{3}] }

\def\redd{$L/L_{\rm Edd}$}
\def\rosat{{\it ROSAT}}
\def\xmm{{\it XMM-Newton}}

\def\kms{$\rm km\,s^{-1}$}

\def\ulum{${\rm ergs\,s^{-1}}$}

\def\unh{${\rm cm^{-2}}$}

\slugcomment{To be appear in {\it The Astrophysical Journal}}
\shorttitle{X-ray Properties of NLS1 with small linewidths}
\shortauthors{Ai et al.}

\begin{document}

\title{X-RAY PROPERTIES OF NARROW-LINE SEYFERT\,1 GALAXIES WITH VERY SMALL BROAD-LINE WIDTHS}

\author{
Y. L. Ai\altaffilmark{1,2},
W. Yuan\altaffilmark{3,1,2},
H. Y. Zhou\altaffilmark{4},
T. G. Wang\altaffilmark{4},
S. H. Zhang\altaffilmark{4}}

\altaffiltext{1}{National Astronomical Observatories/Yunnan
Observatory, Chinese Academy of Sciences, Kunming, Yunnan, P.O. BOX
110, China, ayl@ynao.ac.cn}

\altaffiltext{2}{Key Laboratory for the Structure and Evolution of Celestial
Objects, Chinese Academy of Sciences, Kunming, China}

\altaffiltext{3}{National Astronomical Observatories, Chinese Academy of Sciences,
Beijing, 100012, China, wmy@nao.cas.cn (wmy@ynao.ac.cn)}

\altaffiltext{4}{Key Laboratory for Research in Galaxies and Cosmology,
Center for Astrophysics, University of Science and
Technology of China, Hefei, Anhui, China}

\email{ayl@ynao.ac.cn, wmy@nao.cas.cn}

\begin{abstract}
Narrow-line Seyfert\,1 galaxies (NLS1s) with very small broad-line
widths (say, FWHM(\hb) $\la $ 1200\,\kms)
represent the extreme type of Seyfert\,1 galaxies that
have small black hole masses (\mbh) and/or
high Eddington ratios (\redd).
Here we study the X-ray properties of a homogeneously and optically
selected sample of 13 such objects, termed as
very narrow line Seyfert\,1 galaxies (VNLS1s),
using archival \xmm\ data.
It is found that the Fe K$\alpha$ emission line is at most weak in these objects.
A soft X-ray excess is ubiquitous,
with the thermal temperatures falling within a strict range of 0.1--0.2\,keV.
Our result highlights
the puzzling independence  of the thermal temperature
by extending the relations to
even smaller FWHM(\hb), i.e., smaller \mbh\ ($\sim 10^6$ \msun) and/or higher \redd.
The excess emission can be modeled by
a range of viable models, though the disk reflection and Comptonization models
generally give somewhat better fits over the smeared
absorption and the $p$-free models.
At the Eddington ratios around unity and above, the X-ray spectral slopes
in the 2--10\,keV band are systematically flatter than
the Risaliti et al.'s predictions of the relationship with \redd\ suggested previously.
Short timescale (1--2 hours) X-ray variability is common,
which, together with the variability amplitude computed for some of the objects,
are supportive of the scenario that NLS1s are indeed AGN with
relatively small \mbh.

\end{abstract}


\section{Introduction}

Type 1 active galactic nuclei (AGNs) are characterized by prominent
broad emission lines in their optical/UV spectra.
The lower end of the line widths
is mostly populated by the so-called
narrow-line Seyfert\,1 galaxies (NLS1s), defined as
having the broad hydrogen Balmer lines narrower than $\sim$2000\,\kms\
in full width at half maximum (FWHM)
and the relatively weak \oiii\ lines
\citep[][]{oste85,good89}.
NLS1s show some extreme properties among AGNs \citep[see][for a recent review]{komo08}.
Previous studies  have revealed a set of correlations
among AGN optical emission line and X-ray properties---the
so-called eigenvector 1 (EV 1) correlations \citep[][]{boro92},
which is believed to be driven by the Eddington ratio (\redd).
A narrow \hb\ line is generally associated with strong optical \feii\ and
weak \oiii\ emission \citep[][]{good89, vero01}, a steep soft X-ray
spectral slope \citep[][]{boll96, wang96}, and fast
and large amplitude X-ray variability \citep[][]{leig99,
grup04}.
However, these correlations were found based on AGNs with FWHM(\hb) mostly
broader than $\sim$ 1000\,\kms, below which only a small number of objects
were known by then.
One would expect naively, based on the EV1
correlations, that NLS1s with very small width (say, FWHM$\la$1000\,\kms)
would show even extreme properties in X-ray, i.e., even steeper
soft X-ray slopes and even faster and larger-amplitude variability.
We refer to such AGNs as very
narrow-line Seyfert\,1 galaxies (VNLS1s) hereafter in this paper.

It has been recently found that the gas motion in the broad-line region (BLR) is
virialized \citep[][]{pete00a, onke02} and that the BLR size scales
with optical luminosity with an index of  roughly $0.5$
\citep[][]{kasp05, bent06}.
A combination of these properties naturally links FWHM(\hb)
with black hole mass \mbh\ and \redd\ in a way as FWHM$^4
\sim$ \mbh (\redd)$^{-1}$ \citep{mcha06}.
Therefore, a narrower FWHM(\hb) indicates a larger
(\redd)/\mbh\ ratio, provided that the inclination is not
a dominating effect.
This is what NLS1s are commonly thought to be,
as argued extensively in the literatures
\citep[e.g.,][]{mine00, pete00b, sule00},
and should  be even more extreme for VNLS1s as expected.

As such, the X-ray properties of these extreme
black hole accreting systems are of particular
interest, in light of the following considerations.
Firstly, VNLS1s are well suited for investigating
the soft X-ray excess emission commonly detected in Seyfert 1 galaxies and quasars,
whose origins remain controversial.
One motivation is to attempt to link the soft X-ray excess with
the black body emission from accretion disks,
whose maximum temperature would be the highest among AGNs currently known
(since $T_{\rm max}\propto$[(\redd)/\mbh]$^{1/4}\propto (FWHM)^{-1}$),
and might be detectable with the current X-ray satellites.
Interestingly, this attempt was successful
in at least one AGN, RX\,J1633+4718,
that is also a VNLS1 (FWHM(\hb)$\sim900$\,\kms) but radio-loud,
as recently discovered by \citet[][]{yuan10},
though similar cases are extremely rare.
Alternatively, the observed soft X-ray excess can be mimicked by
relativistically blurred line emission of the reflection component
from a highly ionized inner disk, which may be dominant in
high \redd\ systems \citep[][]{fabi02}.
Secondly, there were suggestions that NLS1s
resemble the fastest accreting states (`high' and `very high'
states) of X-ray binaries, in both the X-ray spectra
\citep[e.g.,][]{poun95, midd07} and
X-ray quasi-periodic oscillations (QPOs) \citep[][]{gier08},
and thus VNLS1s are  more suitable for studying such an analogy.
Thirdly, the X-ray spectral and temporal properties of VNLS1s can be
compared to AGNs with  genuine small mass black holes,
say, \mbh$<10^6$\,\msun\
\citep[e.g.,][]{gree07a, dewa08, mini09}. This may provide possible
insight in the black hole masses of VNLS1s and help distinguish
different models of NLS1s \citep[][]{oste85, mine00, sule00}.

Although some VNLS1s have been studied in X-rays individually in the literature,
systematic studies of their ensemble X-ray properties are rare, however,
given the lack of homogeneously selected samples in the past.
Recently, the X-ray properties of small samples of AGNs
with \mbh$\la 10^6$\,\msun with \xmm\ observations
have been presented \citep[e.g.,][]{dewa08, mini09},
among which several objects are in fact VNLS1 considering their
optical spectral properties and the high \redd.
It was found that these VNLS1s are characterized by strong and rapid
X-ray variability and soft X-ray excess emission.
However, more observations for a larger, homogeneously selected
sample are needed to confirm these results.

Using a large NLS1 sample selected from the SDSS,
\citet{zhou06} found that, to one's surprise,
the previously known FWHM(\hb)--\gams\
(soft X-ray photon index) anti-correlation becomes
flattened at FWHM $\sim$ 1200\,\kms.
Though in the small FWHM regime AGNs having flat \gams\ have previously been
noted to exist\footnote{For such AGNs, their deviation from the above relation is
explained as due to their low luminosity (low \redd); see \citet{laor00}.}, the
lack of expected steep soft X-ray slopes is intriguing.
However, enhanced X-ray absorption in VNLS1s may explain such a trend,
which, though seems to be unlikely, cannot be ruled, since
in \citet{zhou06} the \gams\
(estimated from the \rosat\ hardness ratios)
are subject to large uncertainties.
Detailed spectral modeling of the X-ray spectra with higher spectral
resolution and S/N for these objects is needed to confirm this interesting trend.

Motivated by the above considerations, here
we present a study on the X-ray
properties of a sample of NLS1 with extreme linewidth, FWHM $\leq$ 1200\kms,
using data from archival \xmm\ observations.
The sample and data reduction
are described in Section\,2.
The modeling of the \xmm\ spectra are presented
in Section\,3, with a focus on the soft X-ray excess.
The X-ray temperal properties are presented in
Section\,4, followed by the effective optical to X-ray spectral indices.
The results and their implications are discussed in Section\,6,
and are summarized in Section\,7. We assume a cosmology with $H_0$ =
70\,km\,${\rm s^{-1}}$\,Mpc$^{-1}$, $\Omega_{\Lambda}=0.73$, and
$\Omega_{M}=0.27$. All quoted errors correspond to the 90$\%$
confidence level unless specified otherwise.

\section{Sample and X-ray data}
\subsection{X-ray VNLS1 sample}

We select VNLS1s from a large, homogeneous sample of $\sim$2000
NLS1s built by \citet[][hereafter Zhou'06 sample]{zhou06} from the SDSS DR3,
which can be considered as basically optically selected.
We adopt an operational
linewidth cutoff of FWHM(\hb) $\la$ 1200\,\kms\ for VNLS1s,
 in consideration of the fact that below roughly this value the
\gams--FWHM anti-correlation becomes flattened \citep[see figure 17 in][]{zhou06}.
There are 384 NLS1s meeting this criterion.
We match these VNLS1s
with the 2XMM source catalogue \citep[][]{wats07} using a matching
radius of 5$\arcsec$ and then select those
detected in X-rays with at least 200 net source counts.
We consider radio-quiet\footnote{
Having radio-loudness less than 10, defined as the rest frame flux ratio
between the radio 1.4\,GHz and the optical $g$-band
\citep[see][]{zhou06}} objects only, since X-rays from radio-loud
NLS1s may be contaminated by emission from relativistic jets
\citep[][]{zhou07, yuan08, abdo09}.
The above selection results in 13 objects with reasonable signal-to-noise (S/N) ratios,
which form our working
sample of VNLS1s in this study. The sample objects are
listed in Table\,\ref{optical_property}, and the logs of the \xmm\
observations are summarized in Table\,\ref{XMM-Newton log}.
Among the sample, the  \xmm\ data of seven objects are presented here for the first time;
while the \xmm\ spectra of six objects\footnote{ They are
J0107+1408, J1140+0307, J1357+6525 \citep[][]{dewa08, mini09},
J1246+022 \citep[][]{porq04}, J2219+1207 \citep[][]{gall06b} , and
J1415-0030 \citep[][]{fosc04}.} have been presented previously in
various details in the literatures for different aims.
For the purpose of sample study  using
homogeneously derived results, we
also re-analysis the \xmm\ spectra of these objects,
in the same way as for the other objects whose \xmm\ data
are presented for the first time here.

\begin{figure}
\includegraphics{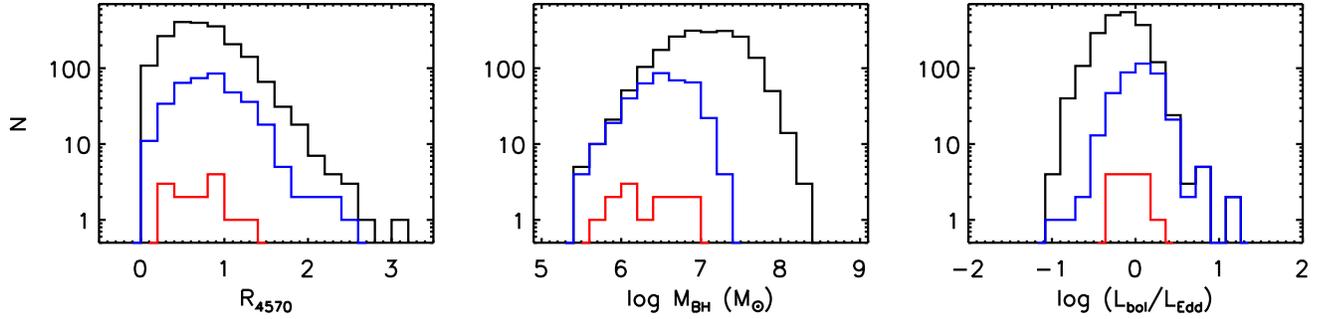}
\caption{Distributions of the \feii\ to H$\beta$ flux ratio (R$_{4570}$), black hole mass and
the Eddington ratio for our sample (red), the total Zhou'06 NLS1s
sample (black) and the whole subsample with FWHM(H$\beta$) $\leq$ 1200\,\kms (blue).
\label{hist_bh_mass}}
\end{figure}

\begin{figure}
\includegraphics{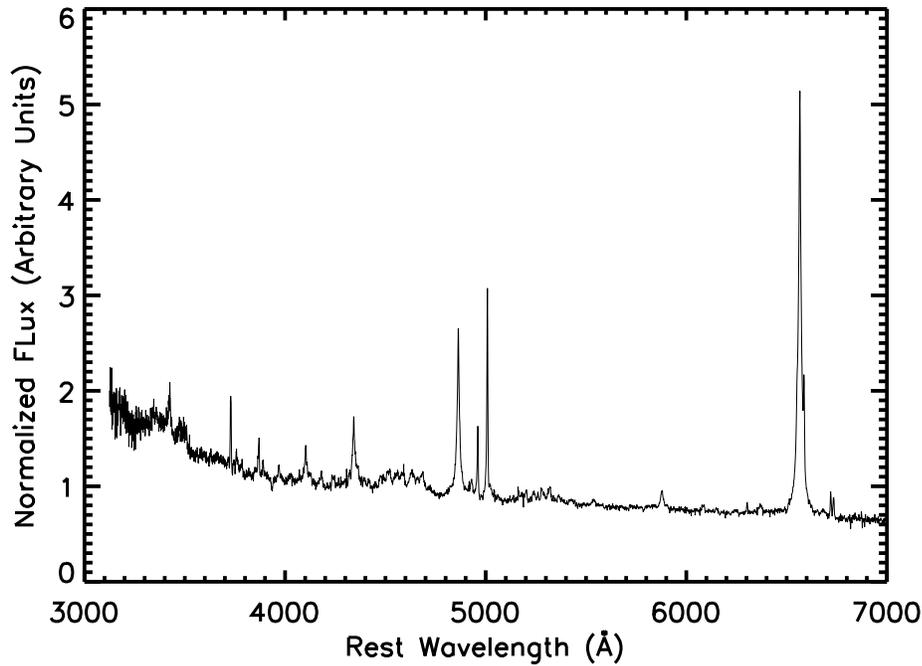}
\caption{Composite optical spectrum of our sample objects
derived from their individual SDSS spectra.
The strong \feii\ multiplet emission is characteristic of typical
NLS1 spectra. \label{composite_spectral}}
\end{figure}

The optical spectral and continuum parameters of the sample objects
are taken from \citet{zhou06} and
given in Table\,\ref{optical_property} .
The black hole masses are estimated from the broad component of the
H$\alpha$ line using the \mbh--linewidth--luminosity relation in
\citet{gree07b}. We also estimate the Eddington ratio \redd~assuming
the bolometric luminosity as 9$\lambda L_{5100}$ \citep{elvi94},
where $\lambda L_{5100}$ is the
monochromatic luminosity at 5100\,\AA. Figure\,\ref{hist_bh_mass}
shows the distributions of several parameters of our working sample,
namely, \mbh,  \redd\ and the optical \feii\
emission multiplets strength $R_{4570}$\footnote{ Defined as the
\feii$(\lambda\lambda4434-4684)$ to \hb\ flux ratio, where
\feii$(\lambda\lambda4434-4684)$ denotes the flux of the \feii\
multiplets integrated over the wavelength range of 4434--4684\,{\rm \AA},
and \hb\ the flux of the broad component of H$\beta$; see
\citet{zhou06}. },
in comparison with those of the
NLS1 sample of Zhou'06 as well as the overall FWHM $< 1200$\,\kms\
(VNLS1) sub-sample. A few remarks can be made concerning the bulk
properties of the sample. Firstly, both of these two VNLS1 sub-samples have
similar $R_{4570}$ distributions to that of the parent sample,
confirming their typical NLS1 nature.
The general strong \feii\ emission is also demonstrated in the
composite SDSS spectrum of our \xmm\ VNLS1 sample (Figure\,\ref{composite_spectral}).
Secondly, our VNLS1s have lower
\mbh\ and higher \redd\ distributions in general than the overall NLS1 sample,
as expected from their narrower linewidths.
Thirdly, our \xmm\ sample is
roughly consistent with the overall VNLS1s
in these distributions. Thus our X-ray VNLS1 sample is not biased
from, but rather representative of, optically selected NLS1s with
the smallest linewidth. This should be kept in mind when comparing
our results with those obtained in previous studies, especially for
X-ray selected NLS1 samples.

\subsection{X-ray observations and data reduction}

The observational data with \xmm\ were retrieved from the \xmm\
science archival center. For all but one object the observations
were operated in the full window mode. For the only exception,
J2219+1207, the MOS cameras were operated in the small window mode,
in which a considerable fraction of the source counts in the wing of
the point spread function (PSF) were lost, and thus only the PN data
are used. The PN observation of J1246+0222 experienced a pile-up,
which is corrected by excising the core of the PSF with a radius of
10\arcsec. Some of the data of the individual cameras as listed in
Table\,\ref{XMM-Newton log} cannot be utilized due to the sources
being  either at the edges or in the gaps of the CCDs, out of the
field of view, or on a bad CCD column.

For \xmm\ data reduction we use the standard Science Analysis System
(SAS, v8.1.0.). The Observation Data Files (ODF)  are processed to
create calibrated events files with `bad' (e.g.,\ `hot', `dead',
`flickering') pixels removed. The time intervals of high flaring
backgrounds contamination are identified and subsequently removed
following the standard SAS procedures and thresholds. Source counts
are extracted from a circle with a radius ranging from, depending on
the source position on the detector, 30\arcsec\ to 65\arcsec~at the
source position, and the background counts from a source-free region
with a usually larger radius. To extract X-ray spectra only X-ray events with the pattern
$\leq 4$ for PN and $\leq 12$ for MOS are used.
Background subtracted light curves are also extracted from cleaned
events files and are subsequently corrected for instrumental effects
(such as vignetting and dead time) using the SAS task `epiclccorr'.

When both available the two MOS spectra of each observation are all
found to be consistent well with each other, and are therefore
co-added to form a single MOS spectrum to increase the S/N (using
the FTOOLS addspec 1.3.0). We set the low-energy cutoff of the spectra
to 0.2\,keV, following the recommendation of the most recent EPIC
calibration status report \citep[][]{guai09}. The high-energy cutoff
is set as such above roughly this energy the background spectrum
starts to dominate. Since there is a diverse range of the spectral
S/N, the high-energy cutoff varies among the sample objects. For
four objects with low spectral S/N, only the 0.2--2.4\,keV range is
used.
For the remaining nine objects, the high-energy cutoff of either 7\,keV or 10\,keV is used,
depending on the spectral quality (see Table\,\ref{XMM-Newton log}).
The EPIC spectra are then grouped in a way that there are at least
25 counts in each energy bin. Spectral fitting is performed using XSPEC
\citep[v.12.3,][]{arna96}. Whenever both PN and MOS spectra are
available for an observation, we perform joint spectral fitting with all
the same spectral parameters tied together except the normalization.

\section{X-ray spectra analysis}
\label{sect:xmodel}

\subsection{Continuum shape and Fe K$\alpha$ emission line}

\begin{figure}
\includegraphics{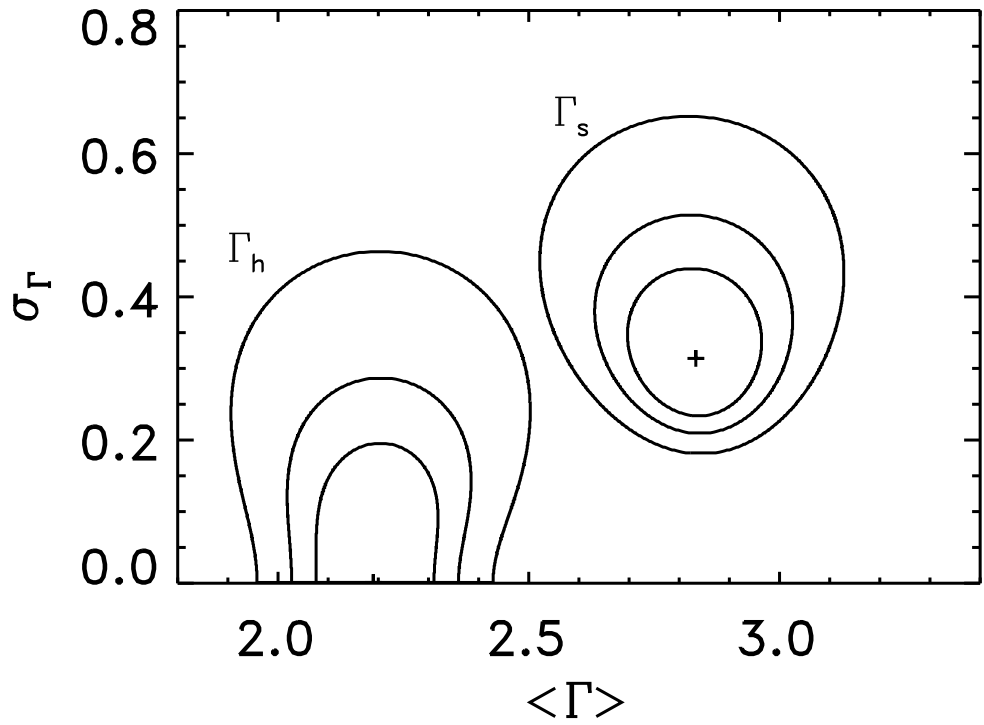}
\caption{Confidence contours (at the
68\%, 90\% and 99\% confidence levels)
of the mean and standard deviation of the intrinsic distributions
(assumed to be Gaussian) of the soft (\gams) and hard (\gamh) X-ray photon indices
for our VNLS1, which are derived using the Maximum-likelihood method
(see text). Pluses indicate the best-estimated values.
\label{gamma_maxi_likh}}
\end{figure}

In order to compare the X-ray continuum slopes with results from other previous AGN studies,
we first characterize the X-ray spectral shape with an absorbed power-law model
in both the soft (0.2--2.4\,keV) and "hard" (2--10\,keV) bands\footnote{For three objects only the 2--7\,keV band is used
since the spectra above 7\,keV are dominated by backgrounds; see
Table\,\ref{XMM-Newton log}.}, respectively.
The results are  listed in Table\,\ref{XMM-Newton fitting}.
In the soft X-ray band,
the model with a neutral absorption column density (\nh) fixed at
the Galactic value (\gnh) yields acceptable fit for about half of the sample objects.
In the remaining objects the fit can be improved by adding
an extra neutral absorber in the objects' rest frame.
The fitted excess absorption \nh\ are small, however, comparable to \gnh.
In only one object, J1415-0030, ionized absorption is required to yield acceptable fit,
with an edge-like feature around 0.6\,keV (see Section\,\ref{sect:soft X-ray excess model}).
We  conclude that intrinsic absorption is not significant in these objects.
We thus suggest that the observed flattening of the
FWHM(\hb)--\gams\ relation below FWHM $\sim$ 1200\,\kms\, as found in \citet[][]{zhou06},
is not caused by X-ray absorption, but most likely an intrinsic property.
For those having more than one measurement, there seems to be little or no
changes in the soft X-ray spectral shape, and the mean \gams\ are calculated.
The fitted \gams\ values range from 2.03 to 3.72.
We quantify the intrinsic  distribution of
\gams\ (assumed to be Gaussian) that is disentangled from
measurement errors using the maximum-likelihood method
as first applied by \citet{macc88};
we find a mean
$\langle$\gams$\rangle=2.83^{+0.19}_{-0.20}$, and a
standard deviation $\sigma=0.31^{+0.20}_{-0.10}$  (90\% confidence), whose confidence
contours are shown in Figure\,\ref{gamma_maxi_likh}.

In the hard X-ray band there are nine objects
having high enough spectral S/N ratios for measuring
photon indices \gamh.
The absorption \nh\ is fixed at the Galactic value.
A power-law
can well reproduce the observed hard X-ray  spectrum in five objects,
whereas the remaining four objects (J0922+5120, J1140+0307,
J1246+0222, and J2219+1207) show a possible broad excess emission feature in the
residuals around 5\,keV.
The fitted  \gamh\ values are in the range of 1.95--2.39.
The maximum-likelihood intrinsic distribution  of \gamh\ has a mean of
$\langle$\gamh$\rangle = 2.19^{+0.19}_{-0.18}$, which is flat for typical NLS1s,
and a small intrinsic scatter, $\sigma =0.0^{+0.29}_{-0.0}$
(see Figure\,\ref{gamma_maxi_likh} for their confidence contours).

No Fe K$\alpha$ emission line feature appears to be present in all
the objects except J1357+6525, which shows a marginal feature of
a narrow-line at around 6.4\,keV. For this object adding a Gaussian
(to an absorbed power-law model in the 2--10\,keV band) improved
slightly the fit, though with only \delchisq=5 for 3 degrees of
freedom (dof), i.e.\ the addition of an extra Gaussian component is
not statistically significant. Thus only upper limits can be derived
on the equivalent width (EW) of any potentially narrow Fe K$\alpha$ line at  6.4\,keV (assuming
$\sigma =$ 10\,eV). The derived upper
limits at the 90\% significance are given in Table\,\ref{XMM-Newton
fitting}. We compared our objects with the Fe K$\alpha$ line EW and
X-ray luminosity relation for AGNs as given in \citet{page04}, and
found that the derived line EW limits are well consistent with the relation.

\subsection{Soft X-ray excess}
\label{sect:soft X-ray excess model}

\begin{figure}
\includegraphics{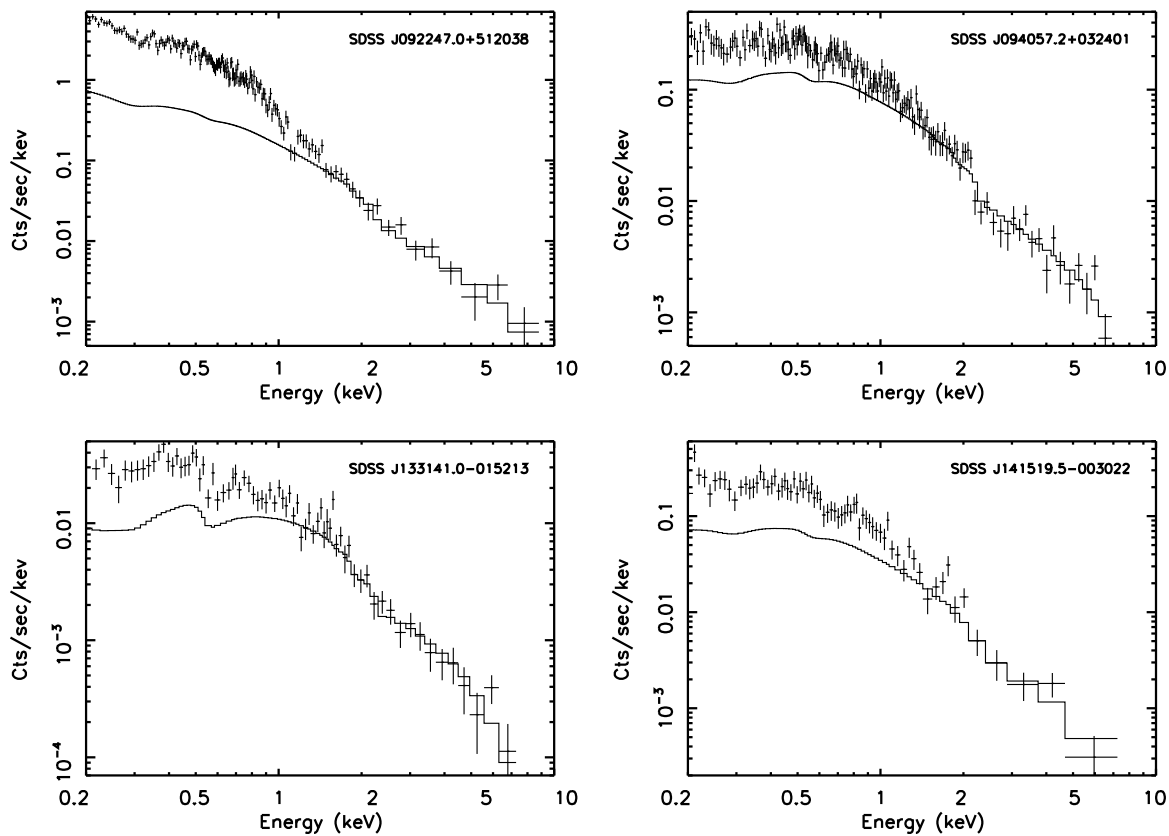}
\caption{\xmm\ spectra of the four objects among our VNLS1 sample (PN spectra except for J1331-0152 of which MOS
spectra are used). The power-law model fitted to
2--10\,keV spectra and its extrapolation to the soft X-ray band is also shown.
Soft X-ray excess emission is clearly present.
\label{hard_soft_ratio}}
\end{figure}

A comparison of the soft and hard X-ray spectral indices obtained above
indicates an overall spectral steepening toward low energies in {\em all} of the objects,
suggesting the presence of the soft X-ray excess.
As demonstration, we show in Figure\,\ref{hard_soft_ratio}
the X-ray spectra of the 4 VNLS1s which are presented for the first time,
and the extrapolation  down to 0.2\,keV of the
power-law model fitted in the hard X-ray band.
Significant excess emission in
the soft X-ray band is prominent, similar to that reported in the
other objects of the sample \citep[e.g.,][]{mini09}, which is also
confirmed here. We thus conclude that the apparent soft X-ray excess
emission is ubiquitous in our  VNLS1 sample.

Given the small \mbh\ and high accretion rate (\redd) in VNLS1s,
the expected blackbody emission from accretion disks
is shifted toward higher energies compared to classical AGNs
with more massive black holes,
and the high energy turnover may start to emerge in the soft X-rays
(e.g., kT$_{max}$$\sim$72\,eV for J0940+0324 assuming a Schwarzschild black hole, e.g.
Peterson 1997).
Thus the blackbody emission directly from the disks might be detected.
We first model the soft X-ray excess with a blackbody.
The model yields
acceptable or marginally acceptable fits for all except for two objects.
For J0107+1408 an additional neutral absorber  is required to improve the fit.
For J1415-0030 a moderately ionized  absorber is needed;
adding an absorption edge
improves the fit significantly with $\Delta\chi^{2}$ = 12 for 3 additional free parameters.
The fitted edge energy is 0.67$\pm$0.03\,keV with an optical
depth of 0.55$\pm$0.2, corresponding to the K-shell binding energy
of ionized Oxygen ions.

\begin{figure}
\includegraphics{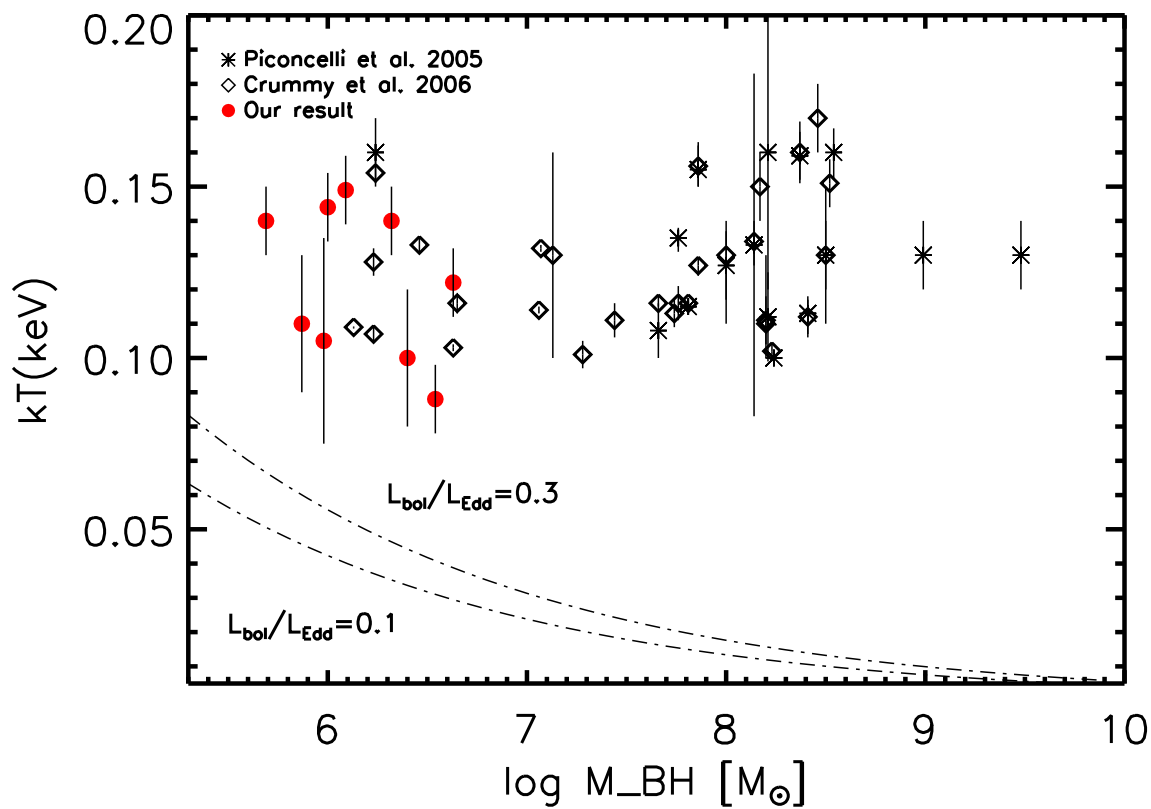}
\caption{The observed temperature of the soft excess is plotted versus the black hole mass. Filled circles are our results, stars for the radio-quiet PG quasars \citep[][]{pico05} and diamonds for type\,1 AGNs \citep[][]{crum06}. The dotted-dashed lines are the maximum temperature expected from the accretion disc.
\label{temp_mass}}
\end{figure}

The fitted temperatures are in the range of 100--200\,eV
(Table\,\ref{model_fitting}),
in agreement with those found in  AGNs having more massive black holes of $10^{7-9}$\,\msun
\citep[][]{gier04, porq04, crum06, bian09}.
Hence our results confirm the extension of
the canonical  100--200\,eV temperatures
down to AGNs with \mbh\ as low as around $\sim10^6$\,\msun\ \citep{mini09}
by adding more objects in this \mbh\ range.
The result is clearly demonstrated in Figure\,\ref{temp_mass},
in which our results are compared with those of AGNs and quasars
with more massive black holes \citep[][]{pico05,crum06}.
These values are still systematically
higher than the maximum temperatures predicted  for standard accretion disks.
The independence of the thermal temperature on \mbh\ over such a wide \mbh\ range
argues against the direct blackbody emission from accretion disks as the origin
of the observed  soft X-ray excess for the vast majority of AGNs,
except for RX\,J1633+4718 \citep{yuan10}.

\begin{figure}
\includegraphics{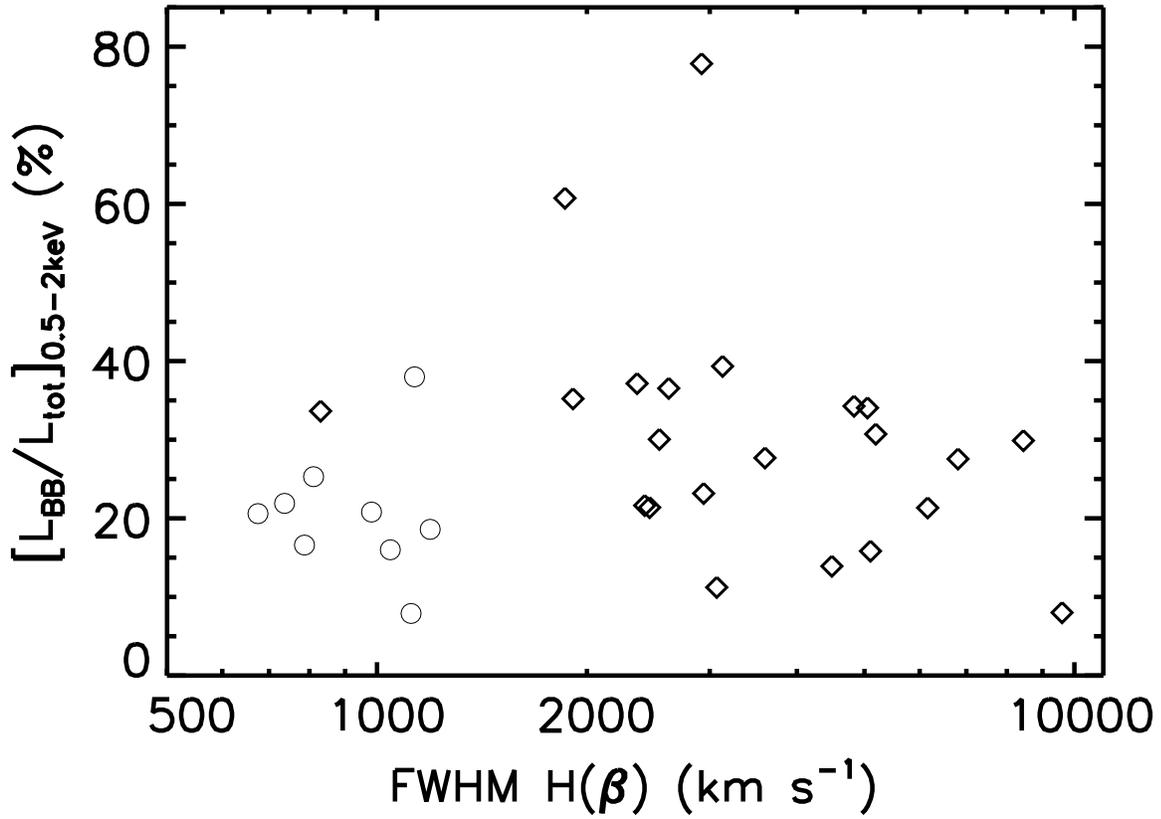}
\caption{Soft X-ray excess strength, parameterized as the ratio of
the blackbody to the total luminosity in the 0.5--2\,keV range,
versus the linewidth for the VNLS1s in our sample (open circles) and
the radio-quiet PG quasars (diamonds) in \citet{pico05}.
\label{soft_str_fwhm}}
\end{figure}

Recent studies suggest that, similar to the blackbody temperature,
the relative strength of the soft X-ray excess also falls within a
relatively small range for PG quasars \citep{pico05}
and Seyfert\,1 AGNs \citep[][]{midd07}
and AGNs with small masses \citep[][]{mini09}.
Here we examine this quantity for the VNLS1s of our sample.
We estimate the relative strength as the
luminosity ratio of the excess component, modeled as a blackbody, to
the total luminosity in the 0.5--2\,keV band.
The values are in the range of 8\%--38\% with a mean of 21\%
(Table\,\ref{model_fitting}).
Apparently, when combined with previous results,
as shown in Figure\,\ref{soft_str_fwhm},
there seems no strong dependence on  \hb\ linewidth over a
large range, FWHM(\hb) = 600--10,000\,\kms\
(a Spearman correlation test probability  of 0.79).
The mean relative strength of our VNLS1s is somewhat smaller than that (31\%) of
the PG quasars in \citep{pico05} derived from their fitting results,
though further confirmation is needed given the relatively small size of our sample.

There are currently several viable models to account for the soft
X-ray excess. Photon trapping in high accreting system where
advection is important \citep[][]{abra88}, or Comptonization of ultraviolet photon from
the accretion disc by electrons as hotter skin
above the disc \citep[][]{czer87,wand88,shim93,czer03}, can
explain the required higher temperature. On the
other hand, absorption or emission lines arising from atomic
processes, when blurred due to relativistic motion, can mimic the
soft X-ray excess. For example, strong relativistically-blurred
emission/absorption lines between $\sim$0.7-2\,keV due to OVII/OVIII and
Fe transition can be produced from ionized disc illuminated by an
underlying hard X-ray continuum (reflection) in the vicinity of the
central black hole \citep{ross93}, or from disc winds \citep{gier04}.
Below we investigate these models by fitting them to the soft X-ray excess spectra.
In the spectral fitting, a power-law continuum
modified by neutral absorption with \nh\ fixed at the Galactic value
is always included.
For objects whose data have been analyzed previously,
we compare our results with previous results individually in Appendix\,A.

\subsubsection{Comptonization}

\begin{figure}
\includegraphics{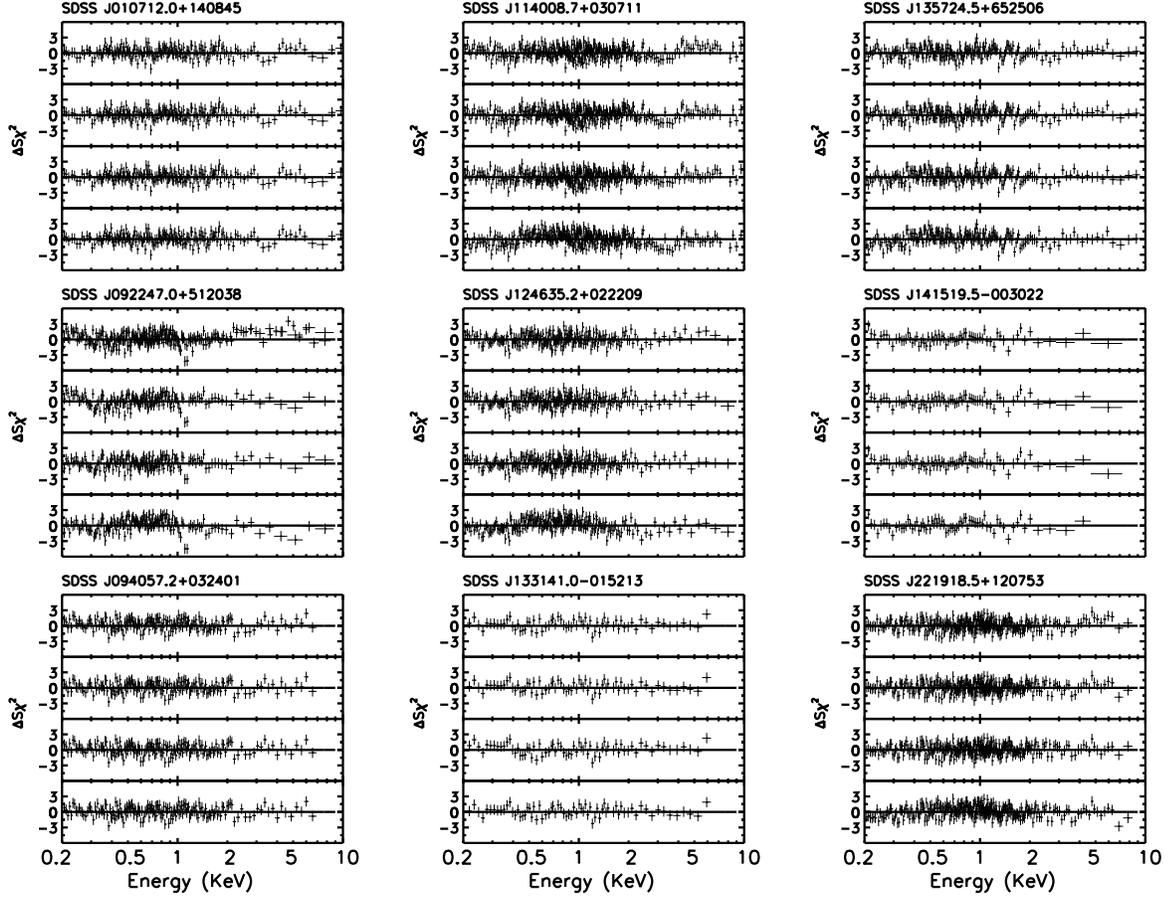}
\caption{Residuals of spectral fits to the \xmm\ spectra with various models to
account for the soft X-ray excess, which are, from top to bottom
for each panel, blackbody, Comptonization, disk reflection,
and smeared absorption model.
An additional blackbody component is added in the disk reflection model
for J0107+1408, J0922+5120 and J2219+1207.
\label{fit_residual}}
\end{figure}

We use the Comptonization model \citep[{\it comptt} in XSPEC,][]{tita94} and
fix the input seed photon energy at the innermost temperature of the
standard accretion disk based on the estimation of the black hole
masses and accretion rates (\redd). In this case the emergent
spectral shape depends on only two parameters, the temperature and
the optical depth of scattering electrons. This model gives
significantly improved fits over, or at least as good as, the above
blackbody fits for all of the objects (see
Table\,\ref{model_fitting}, Figure\,\ref{fit_residual}). The inferred electron temperatures are
found in a relatively small range, $kT_{\rm plasma}$ $\sim$
0.17--0.30\,keV, and the optical depth, $\tau \sim$ 10--25.

\subsubsection{disk reflection}

We use the latest ionized disk reflection model from \citet{ross05} ({\it
reflion} in XSPEC) and in the fits the photon indices of the
ionizing continuum and the observed continuum are tied together, and
the solar abundance is assumed.
For relativistic blurring the laor
kernel model \citep[{\it kdblur} in XSPEC,][]{laor91} is used with
an outer radius fixed at 400\,r$_g$, an emissivity index of the disk
fixed to the standard value of 3, and the inner disk radius allowed
to vary.
Although this model can reproduce acceptable fitting results for most of the
objects,  in three objects the residuals in the soft X-rays
indicate possible contribution from another component.
Following
\citet{mini09}, we then include an additional  blackbody component
in the fits to account for possible contribution from the accretion
disk. This improves the fits for all of the three, namely J0107+1408
($\Delta\chi^{2}/dof$ = 14/3), J0922+5120 ($\Delta\chi^{2}/dof$ =
204/3), and J2219+1207 ($\Delta\chi^{2}/dof$ = 24/2), with the
addition of the blackbody component is statistically significant (a
probability level $<$0.05 using the F-test). The inferred
temperatures are 0.05$\pm$0.008\,keV, 0.06$\pm$0.02\,keV, and
0.08$^{+0.01}_{-0.03}$\,keV, respectively, broadly consistent with
the maximum temperatures at near the inner disk predicted from the
estimated black hole masses and accretion rates.

The disk reflection model, either with or without additional blackbody emission,
provides acceptable fits for all and the best fits for some of the
objects (see Figure\,\ref{fit_residual} and
Table\,\ref{model_fitting}). The best-fit disk inner radius is less than 6r$_{g}$, suggesting a highly spinning
Kerr black hole for most of our objects. The inferred disk
inclination varies from 0$\arcdeg$ to 50$\arcdeg$.
The ionization parameters are $\log\xi
\sim$ 3.13--3.83. We define a parameter `flux fraction'
as the flux ratio of the reflected component to the total component
over the range 0.2--10\,keV \citep[following][]{crum06}. As can be
seen in Table\,\ref{model_fitting}, in most of the objects where
this model gives a good fit, the reflection component dominates
largely the total observed flux in the 0.2--10\,keV band.

\subsubsection{smeared absorption model}

Finally we fit the spectra with the relativistically smeared
absorption model ({\it swind1} in XSPEC). This model provides
acceptable fits for nearly half of the objects but not for the
remaining objects (Table\,\ref{model_fitting}, Figure\,\ref{fit_residual}). The inferred column
densities are in the range of \nh\ $\sim$
0.9--5$\times10^{23}$\,\unh, the ionization parameters $\log\xi
\sim$ 2.75--3.59, and the smearing terminal velocities are very
high, close to 0.5\,c.

\subsubsection{p-free model}

In this work we also try the p-free disk model ({\it
diskpbb} in XSPEC) to account for the soft excess. For some of the
objects, this model gives as good fits as the Comptonization model,
with the inferred temperatures at the inner disk radius of
0.15--0.35\,keV, and the index of the temperature profile $p$ =
0.32--0.64. However, the innermost disk radii derived from the
fitted normalization are significantly less than the estimated
gravitational radii, even after the correction for the spectral
hardening factor ($\sim1.9$ as recommended by \citet{kawa03} for
AGNs with Eddington ratios around 1). We thus consider the $p$-free
model to be physically unrealistic and hence no fitted parameters
are listed here.

\subsubsection{Summary of soft X-ray excess modeling}

In general, the soft X-ray excess of these VNLS1s can be
reproduced by more than one model, which often cannot be
distinguished based on fitting statistics.
However, disk reflection and/or Comptonization are much more preferred
than the other two ones, i.e., smeared absorption model
with marginally improved fitting and p-free model which is physically unrealistic. In the modeling the X-ray
spectra, no additional intrinsic absorption
are required for all the objects except J0107+1408 and J1415-0030, of which additional neural and
warm absorption needed, respectively.

For SDSS J0107+1408,  neutral
absorption with \nh$\sim$6.55--11.41$\times 10^{20}$\,\unh\ is required.
For J1415-0030, in the above spectral fittings for this
object with various soft X-ray excess models, we add ionized
absorption ({\it zxipcf} in XSPEC) applicable to all the emission components. The overall
spectrum and the edge feature can be well reproduced and all the
fits are improved significantly, with a decrease of \delchisq\ = 10
in general. The covering fraction of the absorber is close to
unity, the absorbing column density is in the range of
9.3--15.8$\times10^{20}$\,\unh\ and the ionization parameter of
$10^{0.21-0.38}$, depending on the exact model for the soft
X-ray excess.

\subsection{\gams--FWHM relation}
\begin{figure}
\includegraphics{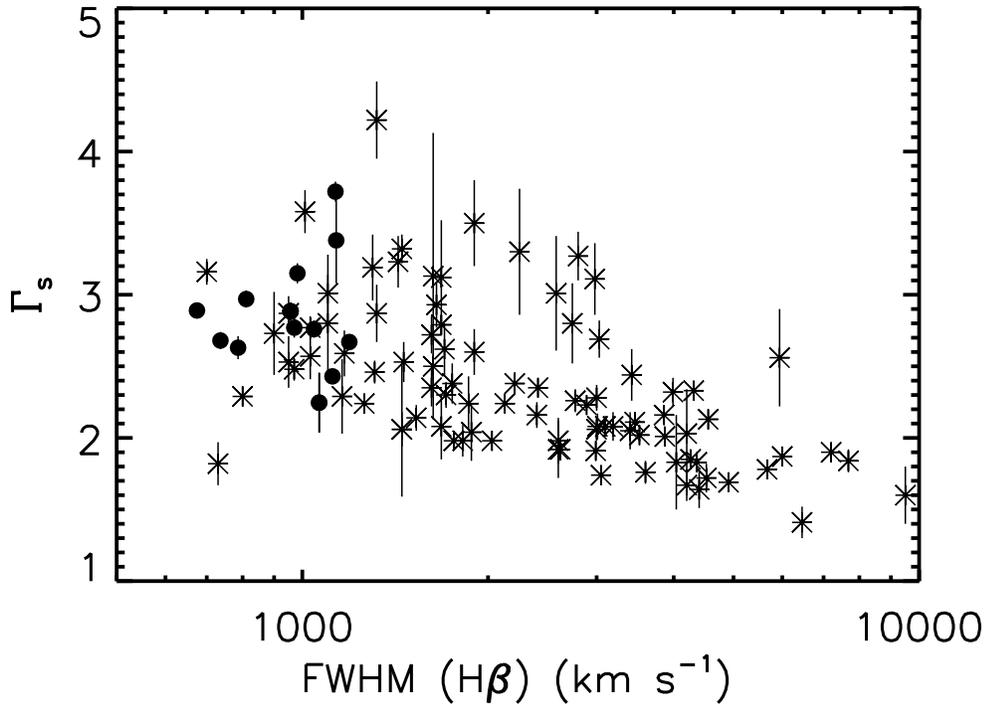}
\caption{Soft X-ray photon index \gams\ versus \hb\
linewidth relation for our VNLS1s (filled circles). Stars are
the results from \citet{grup10}.\label{gamma_fwhm_soft_grup10}}
\end{figure}

We have showed above that there is little or no significant
intrinsic absorption in the X-ray spectra of most of these VNLS1s.
Therefore the photon indices derived in \citet{zhou06}
from the \rosat\ hardness ratios are mostly reliable,
and hence the spectral flattening towards the lower FWHM end \citep{zhou06}
is likely real,
rather than being caused by X-ray absorption.
Recently, \citet{grup10} studied the spectral indices of a sample
of soft X-ray selected AGNs measured with the {\it Swift} XRT,
some of which also have linewidths similar to ours.
We compare
the soft X-ray photon indices of our VNLS1s with
those of the \citet{grup10} sample,
as shown in Figure\,\ref{gamma_fwhm_soft_grup10}.
It can be seen that in the lowest linewidth regime our \gams\ values
are statistically compatible with the result of \citet[][]{grup10} \footnote{Note that the \gams\ values in \citet[][]{grup10} are measured in the 0.2-2\,keV band,
slightly different from 0.2-2.4\,keV adopted here; however, we find from spectral fits that
the differences in \gams\ thus caused are negligible for our objects.}.
It also appears that there is a lack of AGNs having both narrow
FWHM ($\lesssim 1000$\,\kms) and very steep soft X-ray spectra
($\Gamma_{\rm s}\gtrsim 3.5$).
This is consistent with the result suggested by \citet{zhou06}.
Using the combined data points in Figure\,\ref{gamma_fwhm_soft_grup10},
we test explicitly whether there exists a significant flattening
 in the \gams--FWHM relation at FWHM$\lesssim 1000$\,\kms\ using various methods;
the result is inconclusive in the statistical sense, however.
 This might be partly due to the
relative small sample size  and/or the heterogeneity of the combined samples.
A larger and homogeneously selected sample is
needed to test the possible deviation of this well known \gams--FWHM relation
at the low-FWHM end in the future.

\section{X-ray variability}

\begin{figure}
\includegraphics{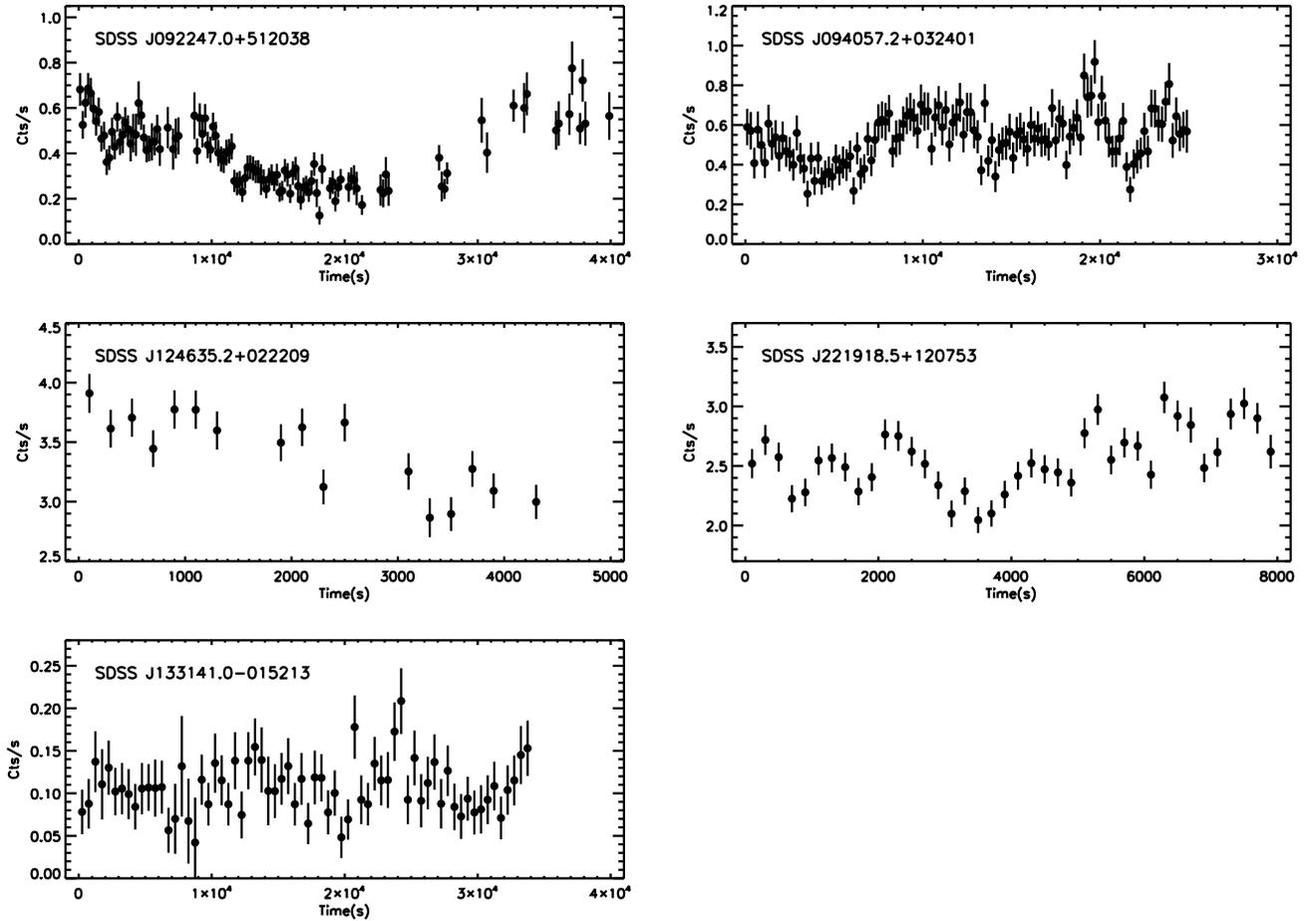}
\caption{\xmm\ X-ray lightcurves in the 0.2--10\,keV band for five
of the VNLS1s in our sample. The time binsize is 200\,s, except for
SDSS J133141.0-015213 (500\,s). \label{lightcurves}}
\end{figure}

Figure\,\ref{lightcurves} shows, as examples, the 0.2-10\,keV lightcurves
for the five objects in our sample, which are  presented
for the first time\footnote{For J0107+1408, J1140-0307, and
J1357+6525 the X-ray lightcurves have been presented in
\citet{mini09} and \citet[][]{dewa08}; for J1415-0030 it was in
\citet{fosc04}.}.
In fact, we find that, for most of the sample objects,
the X-ray flux varied by more than a
factor of 2 on timescales of 1--2 hours within the observational
intervals. We conclude that short timescale variability is common to
VNLS1s. Remarkable flux variations in short-timescales are
evident. We also investigate possible spectral variability within
the observational intervals, which are divided into time bins, using
the hardness ratios; however, no conclusive remarks can be made
mainly due to the insufficient S/N of the data for such a purpose.

The X-ray variability amplitude can be characterized by the excess
variance\footnote{Defined as variance of the lightcurve in a
specific time series after subtracting the contribution expected
from measurement errors \citep[e.g.,][]{vaug99}.}, which has been
found to be strongly correlated with the black hole masses
\citep[][]{lu01,papa04,onei05}. Using the \xmm\ data, some are included in the
current sample, \citet{mini09} extended this relation to AGNs with black hole masses < 10$^6$\,\msun (three included in our sample) and demonstrated that the relationship is relatively tight.
This result indicates that black hole mass is a primary parameter
that drives the relative X-ray variability in AGNs. In the same way as in \citet{mini09}, we
calculate the excess variance for the four objects not presented
previously, that makes use of
lightcurve segments of equal duration (20\,ks) and equal time bin
size (500\,s) in the 2--10\,keV band. However, for only one object,
J0922+5120, the lightcurve has  S/N high enough to yield reliably
determined excess variance, log$\sigma^{2}_{\rm NXS} = -1.15\pm0.68$. We
locate this object (\mbh\ $=10^{6.63}$\msun) on the excess
variance vs.\ black hole mass relation presented in Figure\,8 of
\citet{mini09}, and find that it does follow closely the relation.
It should be noted that, J1140+0307, one of the three objects in \citet{mini09} that are in
our sample is also typical NLS1. As is generally believed that
\mbh\ is the underlying physical parameter that drives the
dependence of the X-ray variability, the fact that these NLS1s
follow the same excess variance--\mbh\ relation as for normal broad-line Seyfert\,1 galaxies (BLS1s)
and quasars tends to validate their \mbh\ estimation. That is to
say, the black hole masses of the VNLS1s in our sample are indeed
relatively small, i.e.\ their narrow Balmer linewidth is caused primarily by
relatively small \mbh, rather than  by a face-on flattened BLR as
claimed in some papers in the literatures.

\section{Optical/UV to X-ray spectral index}

\begin{figure}
\includegraphics{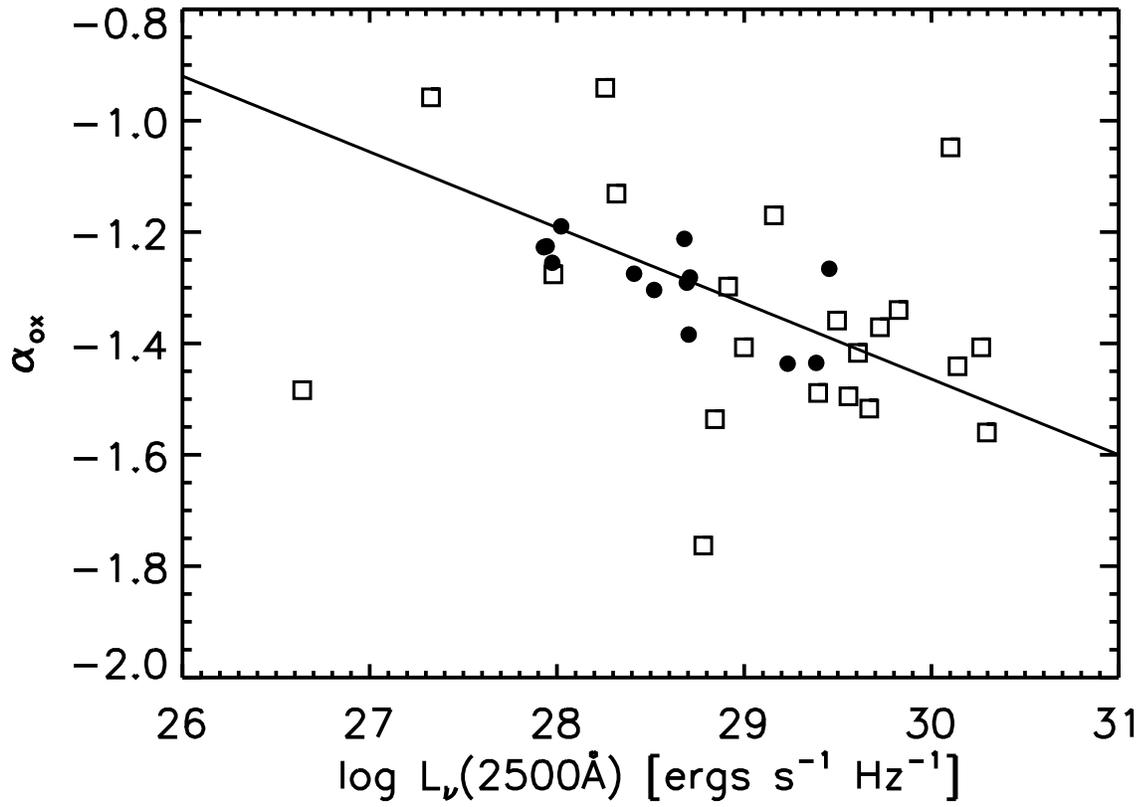}
\caption{Dependence of \alphaox\ on the 2500\,\AA\ monochromatic
luminosity for the VNLS1s in our sample (filled dots) and the
ordinary NLS1s (open squares) in the sample of \citet{gall06a}. The
solid line represents the relation for radio-quiet type\,1 AGNs
given in \citet{stra05}. \label{alpha_ox_2500_lum}}
\end{figure}

The broad-band spectral energy distribution of AGN is commonly
parameterized by the \alphaox\ parameter, \alphaox\ $= -0.3838~\rm
log[L_{\nu}(2500\,\mbox{\AA})/L_{\nu}(2\rm keV)]$ \citep[][]{tana79}, which is
claimed to be correlated with the optical-UV luminosity for Seyferts
and quasars \citep[][]{vign03,yuan98,stra05}. \citet{gall06a}
reported that the objects in their  NLS1 sample, which have
systematically broader linewidths than ours, also follow the same
relation for BLS1s. Here we investigate this relation for our
VNLS1s. The flux density at 2500\,{\rm \AA} is calculated from the
SDSS $u$ band (effective observed-frame wavelength of 3543\,{\rm
\AA}) PSF-magnitude adopting the spectral slope of the composite
SDSS quasar spectrum ($\alpha_{\nu} = -0.44$; Vanden Berk et al.
2001). The \alphaox\ values of our objects are found to lie in the
range from -1.41 to -1.18. We show in
Figure\,\ref{alpha_ox_2500_lum} the \alphaox\ vs. 2500\,{\rm \AA}
monochromatic luminosity relation for our objects, along with the
normal NLS1s from the \citet{gall06a} sample, as well as the
regression relation for normal Seyferts and quasars
\citep[][]{stra05}. It shows that the VNLS1s do follow closely the
\alphaox--$L_{\rm uv}$ relation defined by BLS1s, and consistent
with the result for NLS1s with broader linewidth (FWHM $\gtrsim
1000$\,\kms).

\section{Discussion}

\subsection{Soft X-ray excess in VNLS1s}
\label{sect:disc_sxe}

The ubiquity of the soft X-ray excess in our VNLS1 sample is interesting,
which may have something to do with the generally high \redd\ in our sample.
The reflection model
has been found to be a good description of
the complexity of the X-ray spectra and spectral variability for
several NLS1s \citep[][]{mini04, crum06, ball08, zogh08}. In
some cases the reflection component is found to dominate the
observed X-ray band, which can be explained either in terms of a
corrugated disk or strong gravitational light bending effects
\citep[][]{fabi02}. \citet{mini09} found that this model reproduce
well the \xmm\ spectra of several AGNs with \mbh $\leq$ 10$^6$\,\msun, three of
which are included in this paper. The spectral fitting for the other
objects in our sample also supports this result.
The small inner radii of
the accretion disk derived  argue for fast rotating black holes in
most of these VNLS1s, which may be a consequence of their fast
accretion process. In a few cases the disk thermal emission is
required, which is not unexpected given the relatively high disk
temperatures for small \mbh; this indicates that the disk reflection
model is self-consistent.

The Comptonization model successful explains the soft excess in normal Seyferts
\citep[][]{gier04} and the spectral variability in RE J1034+396
\citep[][]{midd09}. However, as pointed out by \citet{gier04}, the
derived electron temperatures and the optical depth are both found
in a small range ($kT_{\rm e}\sim 0.1-0.3$\,keV and $\tau\sim
10-20$), which is puzzling and requires fine tuning of the disk
parameters. Our result confirms that this is also the case for VNLS1s.

Although the `p-free' model gives statistically acceptable fitting
results in most of the objects, the inferred radii of the innermost
disks are unphysically smaller than the gravitational radii. We thus
consider the simple `p-free' model to be unfeasible. The smeared
absorption model can also reproduce the observed soft X-ray excess
to some extent; however, it is disfavored since it results in
generally poorer fitting statistics compared to the other models for
our sample. Furthermore, the derived velocities are largely
unconstrained at the extreme value and the ionization parameters are
always found to lie in a narrow range.

\subsection{Correlations of spectral indices}

Our result is supportive of the lack of objects with very
steep soft X-ray slopes and FWHM(\hb)$\la1200$\,\kms\ in the \gams--FWHM diagram;
those objects may be expected simply based on this well-known EV1 correlation.
Here we try to link this relationship with \redd\, which is believed to be
the underlying driver of the EV1 correlations.
The soft X-ray spectral slope is primarily
determined by two factors,
the slope of the underlying (hard X-ray) continuum and
the effect of the soft X-ray excess.
We examine the effect of the former by comparing the soft and hard band
spectral indices of our sample objects, and find that they are strongly correlated
with each other
(a Spearman correlation test probability $P<10^{-5}$).
Thus we conclude that soft X-ray slope \gams\ is largely determined by,
or tracing the underlying continuum slope \gamh.

\begin{figure}
\includegraphics{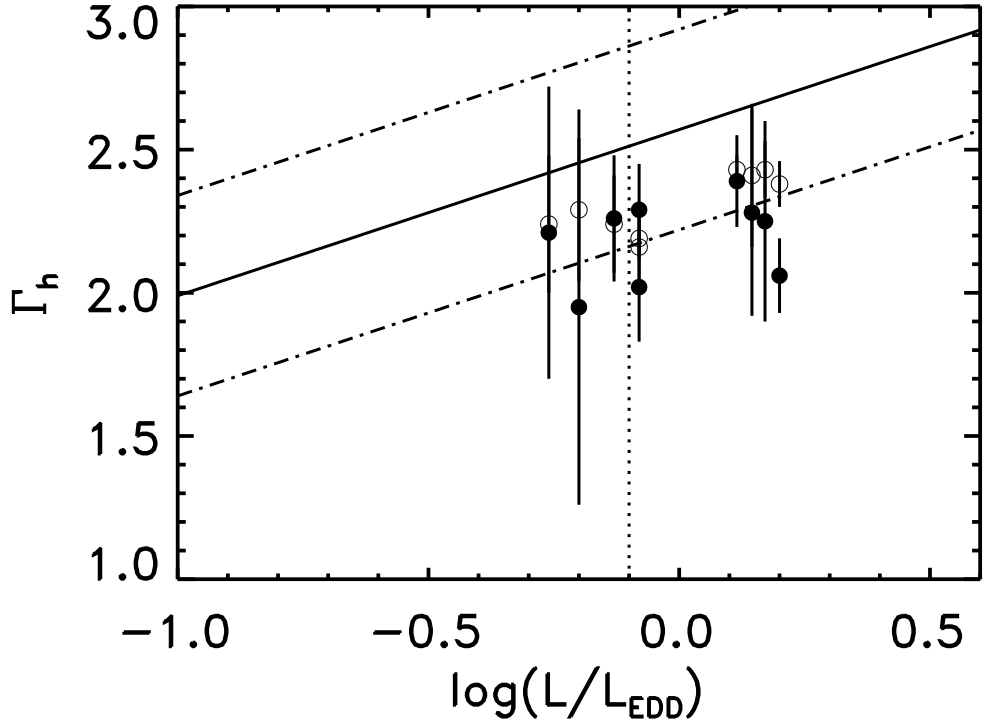}
\caption{Relationship between the hard X-ray (2--10\,keV) photon index
and the Eddington ratio for our sample objects. Filled circles
represent the results from the fits with a simple power-law model
and open circles from a power-law plus disk reflection model. The
solid line is the extrapolation of the relation suggested by
\citet{risa09} (\mbh\ estimated from \hb, the same as in our paper)
and the dash-dotted line represents the dispersion. The vertical
dotted line marks the higher end of the \redd\ range in the Risaliti's sample.
\label{gamma_hard_Edd}}
\end{figure}

Compared to the $\Gamma$--FWHM relation,
a more significant, and perhaps fundamental,
correlation is the \gamh--\redd\ correlation that has been detected
in the range of \redd$<1$ i.e., \citep[][]{shem06,risa09}.
We show  in Figure\,\ref{gamma_hard_Edd}
the \gamh--\redd\ relation for our VNLS1s
whose \gamh\ are available.
Also plotted is, for a comparison, the
regression of \citet{risa09} for the strongest \gamh--\redd\
relation (\mbh\ estimated from \hb, the same as in our paper) and
its extrapolation to the high \redd\ regime
where our sample objects are located.
Interestingly, the \gamh\ values of all our VNLS1s fall
systematically below the extrapolation of the Risaliti et al.'s
relation to high-\redd\ values. To check whether this flattening
might be caused by the enhanced Compton reflection hump---known to
exist in Seyfert galaxies and to make the hard X-ray spectrum
flatten---in high-\redd\ objects, we over-plot the `underlying'
X-ray continuum slope inferred from the above disk reflection model
fitting. As can be seen, the
`underlying' hard X-ray continua (open circles) are indeed slightly
steeper than the `observed' one, but are still systematically
flatter than the prediction of the Risaliti et al.'s relation. We
thus suggest that, at  \redd$\sim 1$ or above, the hard X-ray spectra
become flattened than what the Risaliti et al.'s relation predicts,
at least for the VNLS1s in our sample.
Certainly more observations  are needed to confirm this trend.
If confirmed, this trend in the \gamh--\redd\ relation
may naturally explain the above observed lack of objects with
very steep soft X-ray spectra (i.e. $\Gamma_{\rm s}\gtrsim 3.5$) at the lowest FWHM end.

\subsection{Comparison with IMBH AGNs}

Three objects in our sample were studied by \citet[][]{mini09}
as AGNs with intermediate mass black holes (IMBHs),  a term
sometimes used to refer to black holes with \mbh$<10^{6}$\,\msun\
in the literature \citep[e.g.][]{gree04}.
Since \mbh\ $\propto \rm
FWHM^2$, most IMBH AGNs must have broad line widths falling
within the conventional criterion of NLS1s ($\la 2000$\,\kms),
but may not necessarily possess the characteristics of
typical NLS1s, i.e., strong FeII emission,
high Eddington ratios, and significant soft X-ray excess \citep[e.g.][]{gree04}.
The X-ray properties of IMBH AGN samples have been studied by
several authors \citep[][]{gree07a, dewa08, desr09,mini09}, and a
large diversity has been found. For instance, the soft X-ray
(0.5--2\,keV) photon indices are found to fall into a large range
\citep[\gams\ = 1--2.7,][]{desr09}.
 The flat X-ray spectral slopes,
as well as some other properties, are very similar to those of
typical Seyfert galaxies with \mbh\ = $10^{7-8}$\,\msun.
Some, especially those with low \redd,
do not show soft X-ray excess \citep[][]{iwas00, dewa08},
as in  NGC\,4395, the prototype of this kind.

We suggest that IMBH AGNs,
albeit their small linewidths as for NLS1s,
have diverse observed properties,
depending on the Eddington ratio.
Those accreting at high \redd\ values are probably more NLS1-like,
e.g.\ the presence of a significant soft X-ray excess, strong
FeII, steep \gams, such as the three IMBHs in Miniutti et al.\ (2009)
and also included in our sample as VNLS1.
On the other hand,
there exists a population accreting at low
\redd, which exhibit properties resembling closely those of
classical Seyfert galaxies with more massive \mbh, in both
optical and X-ray
(week FeII, relatively flat \gams, non-ubiquity of the soft X-ray excess).
The observed spectral
properties of Seyfert galaxies depend much more strongly on mass
accretion rate than on black hole mass. In this regards, the
conventional definition of NLS1s may have to be revised. On the
other hand, NLS1s and IMBH AGNs show similar timing property. This
is not surprising given the postulation that the X-ray variability
of AGNs is believed to be largely determined by black hole mass,
rather than accretion rate.

\section{Summary}

NLS1s with very small broad-line widths represent the extreme of
Seyfert\,1 AGNs, which have the largest \redd/\mbh\ ratio among all
AGNs known so far.
Here we investigated the X-ray properties
of a homogeneously selected  sample of NLS1s with FWHM(H$\beta$) $\la$ 1200\,\kms,
using the archival \xmm\ data.
We note that our sample is not complete in the sense that
only those observed with \xmm\ with good spectral S/N ratios are included,
which might be biased toward relatively bright objects in X-rays.
This should be kept in mind when comparing our results with the others.

No significant Fe K$\alpha$ emission line is  detected, which should be
at most weak in such objects.
It is found that
the soft X-ray excess is ubiquitous in the objects which have
the 0.2--10\,keV  spectra available.
The temperatures of this component,
when fitted with a blackbody (or disk blackbody) model,
all fall within 0.1--0.2\,keV, significantly
higher than the prediction of the standard disk model.
Our result highlights the puzzling independence  of the thermal temperature on \mbh\ by
extending it to NLS1s with narrower FWHM(\hb),
i.e., smaller \mbh\ and/or higher \redd.
The failure to ascribe the soft X-ray excess to the Wien tail of the
disk blackbody emission in these VNLS1s
(with similar \mbh\ and \redd\ values to RX\,J1633+4718, though)
highlights the question as to
why RX\,J1633+4718 is so unique \citep[see][for a brief discussion]{yuan10}.
A range of viable models, including Comptonization, disk reflection, smeared
absorption, and the $p$-free model were used to fit the soft X-ray
excess. In general, the disk reflection and Comptonization models
tend to give the best fits.
The relative strength of the soft X-ray excess appears to be
independent of FWHM(\hb) over a large range,
indicating that the excess component is not particularly strong
in these VNLS1s compared to PG quasars with much broader linewidth.

The soft X-ray spectra in 0.2--2.4\,keV have a mean photon index of
\gams=$2.83^{+0.19}_{-0.20}$ with a large intrinsic
 scatter ($\sigma=0.31^{+0.20}_{-0.10}$),
 while the 2--10\,keV spectra have a mean
\gamh\ of $2.19^{+0.19}_{-0.18}$
with a small intrinsic scatter consistent with zero.
Thus VNLS1s have the spectral slopes in both bands no  steeper than
"normal" NLS1s with broader linewidth.
There is little or no intrinsic X-ray absorption in most of these VNLS1s,
indicating that the flattening of the \gams--FWHM anti-correlation below
FWHM$\sim$1200\,\kms, as suggested in \citet[][using a much larger sample
but with \gams\ estimated from hardness ratios]{zhou06},
is not caused by absorption but most likely intrinsic.
Although this trend is not statistically significant when
combining our current sample with the {\it Swift} sample of \citet{grup10},
both with \gams\ derived from spectral fitting,
there appears  a lack of AGNs with both
very narrow FWHM(H$\beta$) ($\la$ 1000\,\kms) and
very steep soft X-ray spectra (i.e., $\Gamma_{\rm s}\gtrsim 3.5$).
Furthermore, there is a tentative hint that the hard
X-ray slopes \gamh\ of our objects fall systematically below the
extrapolation of the suggested \gamh--\redd\ correlation
\citep[][]{risa09} to high \redd\ values.
We argue that these two trends, if confirmed,
might in fact be driven by the same underlying physical process.
Similar to other "normal" NLS1s,
these VNLS1s also follow the same \alphaox--$L_{\rm opt}$ relation
as for normal Seyferts and quasars.

All of the sample objects show rapid variability in X-rays, with two-fold
timescales of 1--2 hours.
The short variability timescales and the conformance
with the variance excess--black hole mass
relation for normal Seyferts tends to suggest
that the black hole masses in these VNLS1s are likely truly small,
as commonly thought, and
the present \mbh\ estimators based on the linewidth--luminosity
scaling relation is applicable to NLS1s.

%%%%%%%%%%%%%%%%%%%%%%%%%%%%%%%%%%%%%%%%%%%%%%%%%%%%%%%%%%%%%%%%%%%%%%%%%%%%%%%%%%%%%%%%%%%
\acknowledgements
We thank the referees for helpful comments and suggestions, which helped to
improve the paper significantly.
Y. Ai is grateful to Stefanie Komossa for her tutoring the X-ray data analysis and
discussion on this work, and her kind hosptality during the visit at MPE.
Y. Ai is grateful to the support by the
MPG--CAS Joint Doctoral Promotion Programme and the hospitality of
MPE, based on which part of the research was carried out.
This work is supported by NSFC grants 10533050, 11033007, the National Basic
Research Program of (973 Program) 2009CB824800, 2007CB815405.
This research is based on observations obtained with \xmm, an ESA science mission
with instruments and contributions directly funded by ESA Memeber
States and NASA.
Funding for the SDSS and SDSS-II was provided by the Alfred P. Sloan Foundation, the
Participating Institutions, the National Science Foundation, the U.S. Department of Energy,
the National Aeronautics and Space Administration, the Japanese Monbukagakusho,
the Max Planck Society,
and the Higher Education Funding Council for England.
The SDSS Web Site is http://www.sdss.org/.

%%%%%%%%%%%%%%%%%%%%%%%%%%%%%%%%%%%%%%%%%%%%%%%%%%%%%%%%%%%%%%%%%%%%%%%%%%%%%%%%%%%%%%%%%%%%%%%%%%%%%%%%%%%%%%%

%\end{document}

\clearpage
%%%%%%%%%%%%%%%%%%%%%%%%%%%%%%%%%%%%%%%%%%%%%%%%%%% table 1, optical properties

\begin{deluxetable}{ccccccccccc}
\tablecolumns{11} \tabletypesize{\scriptsize} \tablewidth{0pt}
\tablecaption{Basic parameters of the sample objects
\label{optical_property}} \tablehead{ \colhead{No.} & \colhead{Name}
&\colhead{SDSS Name} &   \colhead{z}   & \colhead{log$\lambda
L_{\lambda5100}$} & \colhead{FWHM(H$\beta$)} &
\colhead{FWHM(H$\alpha$)}   &   \colhead{F(H$\alpha^{bc}$)}
& \colhead{R$_{4570}$} & \colhead{logM$_{\rm BH}$}  &  \colhead{logL$_{\rm bol}$/L$_{\rm Edd}$} \\
\colhead{(1)}    &   \colhead{(2)}   &   \colhead{(3)}    &
\colhead{(4)}&   \colhead{(5)} & \colhead{(6)}    & \colhead{(7)} &
\colhead{(8)}   & \colhead{(9)} &  \colhead{(10)}   & \colhead{(11)}} \startdata
1&J0107+1408 & J010712.0+140845&0.076&42.96&787$\pm$31&709$\pm$12&1500$\pm$12&0.35$\pm$0.05&5.87&-0.13\\
%2&J0304+0028 & J030417.8+002827&0.044&43.00&1100$\pm$38&841$\pm$17&4339$\pm$38&0.56$\pm$0.04&6.01&-0.23\\
2&J0740+3118 & J074020.2+311841&0.295&44.20&1135$\pm$28&1090$\pm$17&2121$\pm$24&0.92$\pm$0.05&6.90&0.073\\
3&J0922+5120 & J092247.0+512038&0.159&44.01&1132$\pm$27&1002$\pm$13&3058$\pm$23&1.37$\pm$0.04&6.63&0.145\\
4&J0940+0324 & J094057.2+032401&0.060&43.12&1119$\pm$95&810$\pm$34&2249$\pm$36&0.92$\pm$0.09&5.98&-0.08\\
%6&J0952+4937 & J095248.2+493746&0.259&43.80&785$\pm$124&952$\pm$68&365$\pm$15&0.64$\pm$0.25&6.41&0.156\\
%7&J0958+5602 & J095833.9+560225&0.216&43.88&984$\pm$76&1037$\pm$32&1160$\pm$20&0.89$\pm$0.10&6.62&0.034\\
5&J1000+5536 & J100032.2+553631&0.215&43.79&1065$\pm$75&1216$\pm$35&1056$\pm$17&0.28$\pm$0.10&6.76&-0.20\\
6&J1114+5258 & J111443.7+525834&0.079&43.25&970$\pm$60&866$\pm$21&1158$\pm$14&0.65$\pm$0.08&6.04&-0.02\\
7&J1140+0307& J114008.7+030711&0.081&43.12&675$\pm$41&571$\pm$18&1337$\pm$17&0.97$\pm$0.08&5.69&0.200\\
8&J1231+1051& J123126.5+105111&0.304&43.92&957$\pm$23&1200$\pm$332&1143$\pm$255&0.57$\pm$0.05&6.93&-0.23\\
9&J1246+0222& J124635.2+022209&0.048&43.49&811$\pm$27&709$\pm$15&11669$\pm$118&0.83$\pm$0.04&6.09&0.171\\
10&J1331-0152& J133141.0-015213&0.145&43.56&1192$\pm$42&1044$\pm$16&1929$\pm$15&0.35$\pm$0.03&6.54&-0.20\\
11&J1357+6525& J135724.5+652506&0.106&43.14&737$\pm$41&694$\pm$16&1471$\pm$16&0.45$\pm$0.07&6.00&-0.08\\
12&J1415-0030& J141519.5-003022&0.134&43.36&1045$\pm$27&954$\pm$11&1735$\pm$13&0.76$\pm$0.04&6.40&-0.26\\
%16&J1502+4054& J150245.4+405437&0.232&43.96&680$\pm$25&797$\pm$11&1632$\pm$14&1.06$\pm$0.11&6.46&0.270\\
%17&J1633+3713& J163338.3+371314&0.115&43.70&895$\pm$12&843$\pm$6&8410$\pm$38&0.37$\pm$0.02&6.53&-0.05\\
%18&J1654+3925& J165408.2+392533&0.069&43.31&1159$\pm$16&1004$\pm$7&6732$\pm$30&0.58$\pm$0.02&6.43&-0.34\\
13&J2219+1207& J221918.5+120753&0.081&43.66&982$\pm$38&886$\pm$15&4369$\pm$38&1.11$\pm$0.06&6.32&0.115\\
\enddata
\tablecomments{Col.(2): abbreviated name of objects;
Col.(3): SDSS name;
Col.(4): redshift;
Col.(5): monochromatic luminosity at 5100\, {\rm \AA} (ergs\,s$^{-1}$);
Col.(6): H$\beta$ linewidth (km\,s$^{-1}$);
Col.(7): H$\alpha$ linewidth (km\,s$^{-1}$);
Col.(8): H$\alpha$ broad component flux (10$^{-17}$\,ergs\,s$^{-1}$\,cm$^{-2}$);
Col.(9): the optical Fe\,{\footnotesize II} strength relative to the H$\beta$ broad component;
Col.(10): black hole mass (M$_{\odot}$);
Col.(11): Eddington ratio}
\end{deluxetable}

%%%%%%%%%%%%%%%%%%%%%%%%%%%%%%%%%%%%%%%%%%%%%%%%table 2, observation log of XMM
\begin{deluxetable}{ccccccc}
\tablecolumns{7}\tablewidth{0pt}
\tabletypesize{\small}
\tablecaption{Log of \xmm\ observations \label{XMM-Newton log}}
\tablehead{
\colhead{Name}  &    \colhead{Date} &  \colhead{Off-Axis}
&   \colhead{}  &\colhead{Exposure time}  &    \colhead{}&  \colhead{Note}   \\
\cline{4-6}
\colhead{}   &    \colhead{}  &
\colhead{}   & \colhead{PN}   & \colhead{MOS1} &   \colhead{MOS2} &   \colhead{} \\
\colhead{(1)}   &    \colhead{(2)}  &
\colhead{(3)}   & \colhead{(4)}   & \colhead{(5)} &   \colhead{(6)} &   \colhead{(7)} }
\startdata
J0107+1408&  2005-07-22&  0.0&  15.1& 27.1&27.1&  c \\
J0740+3118&  2001-04-19&  12.7&  - &2.2&-&  a \\
J0922+5120& 2005-10-08 &0.0& 5.2 &19.2 &- &c \\
J0940+0324&  2005-10-30&  11.8&  21.9&26.1&26.2&  b\\
J1000+5536& 2001-04-13&12.8&-&-&8.2&   a\\
          &  2003-10-14&  13.5&  14.2&-&-&   a\\
J1114+5258&  2003-04-25&  11.3&  3.9&6.6&7.0&  a \\
J1140+0307&  2005-12-03&  0.0&  30.8& 38.5 & 39.2 &  c \\
J1231+1051&  2003-07-13&  12.0&  -&45.3&-&a\\
          & 2005-12-13 & 12.0& -&-&68.2&a \\
          &  2005-12-17&  12.0&  -&91.6&91.6&  a \\
J1246+0222& 2001-06-17&  0.0&  3.1& - &-&  c \\
J1331-0152& 2001-07-29&  12.0&  -&32.3&32.3&  b \\
J1357+6525& 2005-04-04&  0.0&  14.5&21.1&20.6&  c \\
J1415-0030& 2003-02-08&  10.0&  9.5&13.7&14.2&   b\\
J2219+1207&  2001-06-07&  0.1&  7.2&-&-&   c \\
\enddata
\tablecomments{Col.(2): observation date;
Col.(3): off axis angle in arcmin;
Col.(4)--(6): cleaned exposure time of the three EPIC cameras in kilo-second;
Col.(7): energy range in which source spectrum is extracted, (a)
0.2--2.4\,keV, (b) 0.2--7\,keV, (c) 0.2--10\,keV.}
\end{deluxetable}

%%%%%%%%%%%%%%%%%%%%%%%%%%%%%%%%%%%%%%%%%%%%%%%%%%%%%%%%Table 3, XMM spectral fitting with power-law
\begin{table}
\begin{center}
\caption{\xmm\ spectral fits with an absorbed power-law model
\label{XMM-Newton fitting}} \scriptsize
\begin{tabular}{c|c|cccc|ccc|c}\\
\hline
\hline
Name &     $N_{\rm H}^{\rm Gal}$ &   $N_{\rm H}^{\rm in}$ &
$\Gamma_{\rm soft}$ &   log L$_{\rm soft}$  & $\chi^2_{\nu}$/dof &
$\Gamma_{\rm hard}$  &  log L$_{\rm hard}$ &   $\chi^2_{\nu}$/dof & EW  \\
(1)  &   (2)  &  (3) &   (4)&   (5) &   (6) &   (7) & (8) &   (9) &   (10) \\
\hline
J0107+1408& 3.37& $4.2\pm0.8$ & $2.63\pm0.08$& 43.07& 1.0/230 & 2.26$\pm$0.22 & 42.47& 1.3/44 & $<$45 \\
J0740+3118&  4.32&  -&  $3.38\pm0.30$& 44.72 & c-stat & - & -& -& -\\
J0922+5120& 1.20& 2.7$\pm$0.5& 3.72$\pm$0.07& 44.23 & 1.7/226& 2.28$\pm$0.36& 43.31& 1.3/24 & $<$228 \\
J0940+0324&  3.25&  -&  $2.43\pm0.03$& 43.00& 1.0/244&  2.02$^{+0.19}_{-0.25}$& 42.56 &   1.1/39  & $<$497 \\
J1000+5536&  0.83&  -&  2.03$\pm$0.38& 43.33& c-stat  & - & -& -& -\\
   &  -&  -&  $2.34\pm0.25$& 43.14& 1.5/18  & - & -& -& -\\
J1114+5258&  0.99&  -&  $2.77\pm0.07$& 43.16 & 0.8/60  & - & -& -&  -\\
J1140+0307&  1.88& 1.3$\pm$0.2&  $2.89\pm0.03$& 43.44& 1.3/375&  2.06$\pm$0.13& 42.65& 1.0/94  & $<$111 \\
J1231+1051&  2.40&  -&  $2.89\pm0.18$& 44.10& 0.9/26 & - & -& -& - \\
&  -&  -& 2.80$\pm$0.16& 43.83& 1.5/28 & - & -& -& -\\
&  -&  -&  $2.90\pm0.07$& 43.93&  1.0/69 & - & -& -& -\\
J1246+0222& 1.74& 1.83$\pm$0.5& $2.97\pm0.06$& 43.65& 1.2/198& 2.25$^{+0.35}_{-0.23}$& 42.74& 1.1/17& $<$214 \\
J1331-0152& 2.33&  -&  $2.67\pm0.05$& 43.65& 1.0/58&  1.95$^{+0.69}_{-0.35}$& 43.18 &  0.8/11  & $<$361 \\
J1357+6525& 1.21& $2.3\pm0.4$& $2.68\pm0.06$& 43.43& 1.1/265&  2.29$^{+0.16}_{-0.32}$& 42.85&  1.1/46 & $<$276 \\
J1415-0030&  2.84&  -&  $2.76\pm0.06$& 43.49& 1.0/104&  2.21$^{+0.51}_{-0.47}$& 42.81 &  1.4/11  & $<$842 \\
J2219+1207& 4.81& $2.8\pm0.7$& $3.15\pm0.07$& 44.19& 1.0/264& 2.39$\pm$0.16& 43.21 & 1.1/38 & $<$125 \\
\hline
\end{tabular}
\tablecomments{Col.(2): Galactic column density in $10^{20}$\,\unh;
Col.(3): column density of intrinsic neutral absorption in
the object's rest frame in $10^{20}$\,\unh;
Col.(4): fitted power-law photon index in the soft X-ray band (0.2--2.4\,keV);
Col.(5): absorption corrected luminosity in 0.2-2.4\,keV in \ulum;
Col.(6): reduced $\chi^2$;
Col.(7): fitted power-law photon index in the hard X-ray band
(mostly in 2--10\,keV; see Table\,\ref{XMM-Newton log});
 Col.(8): absorption corrected
luminosity in 2--10\,keV in \ulum;
 Col.(9): reduced $\chi^2$;
Col.(10): rest frame equivalent width of the Fe K$\alpha$
line in units of eV}
\end{center}
\end{table}

%%%%%%%%%%%%%%%%%%%%%%%%%%%%%%%%%%%%%%%%%%%%%%%%%%%%%%%%%%Table 4, soft X-ray excess model fitting
\clearpage
\begin{longtable}{ccccccc}
\caption{Results of spectral fits in 0.2-10\,keV with different models for the soft excess \label{model_fitting}}
\scriptsize
\endfirsthead
\hline
\caption{(continued)}
\endhead
\hline
\hline
\multicolumn{2}{l}{Blackbody} & \\
\hline
Name &   $\Gamma$   & \it{k}T\rm\,(keV)  & BB/total$^{(a)}$&  $\chi^2$/dof  \\
\hline
J0107+1408 &2.39$^{+0.08}_{-0.10}$&  0.11$\pm0.02$& 0.166& 277/264\\
J0922+5120 &3.18$\pm$0.09& 0.12$\pm$0.01& 0.380& 356/255 \\
J0940+0324 &2.29$\pm$0.05&  0.11$^{+0.04}_{-0.03}$& 0.079 &278/273\\
J1140+0307 &2.56$\pm$0.04&  0.14$\pm$0.01& 0.206 & 570/472 \\
J1246+0222 &2.61$\pm$0.02&  0.15$\pm$0.01& 0.253 & 220/212 \\
J1331-0152 &2.27$\pm$0.06&  0.09$\pm$0.01& 0.186 & 61/68  \\
J1357+6525 &2.36$\pm$0.09&  0.14$\pm$0.01& 0.219 & 319/303  \\
J1415-0030 &2.42$^{+0.12}_{-0.20}$ &0.10$\pm$0.02&  0.160 & 102/109 \\
J2219+1207 &2.70$\pm$0.04&  0.14$\pm$0.01 & 0.208 & 293/293  \\
\hline
\multicolumn{2}{l}{Comptonization} & \\
\hline
Name &   $\Gamma$   & \it {k}T$_{plasma}$\rm\,(keV)  & $\tau$  &  $\chi^2$/dof  \\
\hline
J0107+1408& 2.32$\pm0.06$& 0.25$\pm0.07$& 10.9$\pm4.3$& 272/263\\
J0922+5120& 2.09$\pm$0.19& 0.17$\pm$0.01 & 19.9$^{+2.5}_{-0.6}$& 325/254 \\
J0940+0324& 2.04$\pm0.19$& 0.30$\pm0.10$&  13.5$\pm2.1$& 269/272\\
J1140+0307& 2.19$\pm0.15$& 0.22$\pm0.03$&18.7$\pm7.9$& 540/471 \\
J1246+0222& 2.34$\pm0.23$& 0.21$\pm0.02$& 21.0$\pm5.5$& 217/211 \\
J1331-0152& 2.12$\pm0.21$& 0.17$\pm0.07$& 17.1$\pm7.9$& 60/67\\
J1357+6525& 2.22$\pm0.04$& 0.19$\pm0.02$&  23.5$\pm7.0$& 320/302 \\
J1415-0030& 2.04$\pm0.20$& 0.20$\pm0.06$& 20.0$\pm4.8$& 99/108\\
J2219+1207& 2.36$\pm0.15$& 0.21$\pm0.04$& 18.1$\pm7.0$ & 279/292\\
\hline
\multicolumn{2}{l}{Disk Reflection} & \\
\hline
Name &  $\Gamma$ & r$_{in}/r_{g}$ & log$\xi$  & kT\,(keV)$^{(b)}$ & Flux Frac.$^{(c)}$ & $\chi^2$/dof \\
\hline
J0107+1408& 2.24$\pm$0.17 & 4.93$^{+3.31}_{\rm - }$ & 3.45$\pm$0.4 & 0.05$\pm$0.008 & 0.3 & 260/259 \\
J0922+5120& 2.41$\pm$0.25 & 1.73$^{+11.07}_{\rm - }$ & 3.33$^{+0.13}_{-0.12}$ & 0.06$\pm$0.01& 0.5 & 257/250 \\
J0940+0324& 2.16$\pm$0.12 & 1.41$^{+8.98}_{\rm - }$ & 3.76$^{\rm + }_{-0.21}$ & - & 0.9 & 273/271 \\
J1140+0307& 2.38$\pm$0.08 & 3.86$^{+0.89}_{\rm - }$ & 3.83$^{+0.05}_{-0.15}$ & - & 0.9 & 524/470  \\
J1246+0222& 2.43$\pm$0.10 & 3.82$^{+2.32}_{\rm - }$ & 3.75$^{\rm + }_{-0.19}$ & - & 1.0 & 212/210  \\
J1331-0152& 2.29$\pm$0.25 & 1.24$^{+3.15}_{\rm - }$ & 3.27$^{\rm + }_{-0.25}$ & - & 0.3 & 69/66 \\
J1357+6525& 2.19$\pm$0.07 & 6.05$^{+1.91}_{-3.03}$ & 3.28$^{+0.30}_{-0.23}$ & - & 0.5 & 320/301\\
J1415-0030& 2.24$\pm$0.24 & 1.25$^{+2.24}_{\rm - }$ & 3.13$^{+0.12}_{-0.43}$ & - & 0.7 & 104/109 \\
J2219+1207& 2.43$\pm$0.05 & 4.46$^{+2.15}_{\rm - }$ & 3.79$^{\rm + }_{-0.15}$ & 0.08$^{+0.01}_{-0.03}$ & 0.8 & 272/289\\
\hline
\multicolumn{2}{l}{Smeared Absorption} & \\
\hline
Name &  $\Gamma$ &
$\log\xi$  & $N_{H}$ ($10^{22}$\,\unh) & $\sigma$ (v/c) & $\chi^2$/dof \\
\hline
J0107+1408& 2.47$\pm$0.21 & 3.00$\pm$0.71 & 12$_{-6}^{\rm + }$ & 0.50$_{-0.16}^{\rm + }$ & 278/264\\
J0922+5120& 2.69$\pm$0.07& 3.11$\pm$0.04& 50$_{-2}^{\rm + }$ & 0.5$_{-0.05}^{\rm + }$ & 388/255 \\
J0940+0324& 2.22$\pm$0.10 & 3.46$\pm$0.35 & 35$_{-24}^{\rm + }$ & 0.44$_{-0.28}^{\rm + }$ & 275/273\\
J1140+0307& 2.37$\pm$0.03 & 3.40$\pm$0.05 & 50$_{-3}^{\rm + }$ & 0.5$_{-0.05}^{\rm + }$ & 607/472\\
J1246+0222& 2.56$\pm$0.09 & 3.58$\pm$0.10 & 50$_{-8}^{\rm + }$ & 0.47$_{-0.08}^{\rm + }$ & 291/211\\
J1331-0152& 2.42$\pm$0.07 & 2.75$\pm$0.36 & 14$_{-9}^{+25}$ & 0.5$_{-0.19}^{\rm + }$ & 60/67\\
J1357+6525& 2.26$\pm$0.06 & 3.59$\pm$0.06 & 50$_{-9}^{\rm + }$ & 0.5$_{-0.24}^{\rm + }$ & 359/303\\
J1415-0030& 2.60$\pm$0.12 & 3.52$\pm0.50$ & 9$_{-5}^{\rm + }$ & 0.49$_{-0.16}^{\rm + }$ & 112/110\\
J2219+1207& 2.60$\pm$0.20 & 3.43$\pm$0.03 & 50$_{-5}^{\rm + }$ & 0.5$_{-0.05}^{\rm + }$ & 320/292\\
\hline
\end{longtable}
\emph{Notes.}------For SDSS J0107+1408 with free absorption in the fitting.
For SDSS J1415-0030 an additional ionized
absorption is applied (see text). The blank parameter errors denote that the upper or
lower limits are outside of the tabulated parameters range, which are considered to be not
physically meaningful. $^{(a)}$
Luminosity ratio of the blackbody component to the total component
in 0.5--2\,keV. $^{(b)}$Inferred temperatures of an additional
blackbody component. $^{(c)}$Flux ratio of the reflected component
to the total component in 0.2--10\,keV.

\clearpage

%%%%%%%%%%%%%%%%%%%%%%%%%%%%%%%%%%%%%%%%%%%%%%%%%%%%%%%%%%%%%%

%%%%%%%%%%%%%%%%%%%%%%%%%%%%%%%%%%%%%%%%%%%%%%%%%%%%%%%%%%%%

%%%%%%%%%%%%%%%%%%%%%%%%%%%%%%%%%%%%%%%%%%%%%%%%%%%%%%%%%%%%%%

%%%%%%%%%%%%%%%%%%%%%%%%%%%%%%%%%%%%%%%%%%%%%%%%%%%%%%%%%%%%%%%%

%%%%%%%%%%%%%%%%%%%%%%%%%%%%%%%%%%%%%%%%%%%%%%%%%%%%%%%%%%%%%%%%

%%%%%%%%%%%%%%%%%%%%%%%%%%%%%%%%%%%%%%%%%%%%%%%%%%%%%%%%%%%%%%%%

%%%%%%%%%%%%%%%%%%%%%%%%%%%%%%%%%%%%%%%%%%%%%%%%%%%%%%%%%%%%%%%%

%%%%%%%%%%%%%%%%%%%%%%%%%%%%%%%%%%%%%%%%%%%%%%%%%%%%%%%%%%%%%%

%%%%%%%%%%%%%%%%%%%%%%%%%%%%%%%%%%%%%%%%%%%%%%%%%%%%%%%%%%%%%%%%%%%

%%%%%%%%%%%%%%%%%%%%%%%%%%%%%%%%%%%%%%%%%%%%%%%%%%%%%%%%%%%%%%%%%%%%

%%%%%%%%%%%%%%%%%%%%%%%%%%%%%%%%%%%%%%%%%%%%%%%%%%%%%%%%%%%%%%%%%%%%

\clearpage

%%%%%%%%%%%%%%%%%%%%%%%%%%%%%%%%%%%%%%%%%%%%%%%%%%%%%%%%%%%%%%%%%%%%%%%%%%%%%%%%%%%%%%%%%%%
\begin{appendix}

\section{COMPARISONS WITH PREVIOUS RESULTS FOR INDIVIDUAL OBJECTS}
\label{sect:compar}

For six objects in our sample, the \xmm\ data have been presented
previously. Here we compare our results with those in the
literatures.

{\it J0107+1408, J140+0307, J1357+6525} ---
These three objects were studied as IMBH AGNs by  \citet{mini09} and \citet{dewa08}.
For the model fits with
blackbody, ionized disk reflection, and smeared absorption, our results are
consistent with the previous ones except that the disk ionization parameters
of ours are somewhat higher than those in \citet{mini09}.
In addition, we fit the spectra with
the Comptonization and `p-free' model, which was not considered in the previous papers.

{\it J1246+0222} --- The fitting results for the
soft X-ray excess component are consistent with those presented in
the literatures \citep[][]{porq04, crum06, midd07}.
However, in the simple power-law fitting we find a soft X-ray photon
index of 2.97$\pm$0.06 while \citet{porq04} gave the value of 3.72$\pm$0.09.

{\it J1415-0030} --- The \xmm\ data was presented previously by
\citet{fosc04}, who fitted the  spectrum with
the simple power-law plus blackbody model only, and gave a somewhat
flatter photon index (1.8$\pm$0.2) than ours ($2.42^{+0.12}_{-0.20}$).
We find that there are an edge-like feature around 0.6\,keV for J1415-0030.

{\it J2219+1207} --- The spectrum was fitted with a power-law plus
blackbody  model by \citet{gall06b}, giving a result consistent
with ours. The authors also noted the broad excess emission feature
around 5.8\,keV, which was fitted with a broad emission line;
while in our work the disk
reflection model can well reproduce this feature.

\end{appendix}

\end{document}